\documentclass[revtex4]{emulateapj}
\usepackage{natbib}
\usepackage{longtable}
\usepackage{pdflscape}
\usepackage{color} 
\usepackage{morefloats}
\usepackage{gensymb}

\newcommand{\kms}{\ifmmode {\rm km\ s}^{-1} \else km s$^{-1}$\fi}
\newcommand{\ergs}{\ifmmode {\rm erg\ s}^{-1} \else erg s$^{-1}$\fi}
\newcommand{\ergscm}{\ifmmode {\rm erg\ s}^{-1} \else erg s$^{-1}$ cm$^{-2}$\fi}
\newcommand{\Msun}{\ifmmode {\rm M}_{\odot} \else $M_{\odot}$\fi }
\newcommand{\Lsun}{\ifmmode {\rm L}_{\odot} \else L$_{\odot}$\fi}
\newcommand{\qo}{\ifmmode q_{\rm o} \else $q_{\rm o}$\fi}
\newcommand{\Ho}{\ifmmode H_{\rm o} \else $H_{\rm o}$\fi}
\newcommand{\ho}{\ifmmode h_{\rm o} \else $h_{\rm o}$\fi}

\newcommand{\vFWHM}{\ifmmode v_{\mbox{\tiny FWHM}} \else
                    $v_{\mbox{\tiny FWHM}}$\fi}
\newcommand{\CCF}{\ifmmode F_{\it CCF} \else $F_{\it CCF}$\fi}
\newcommand{\ACF}{\ifmmode F_{\it ACF} \else $F_{\it ACF}$\fi}
\newcommand{\Halpha}{\ifmmode {\rm H}\alpha \else H$\alpha$\fi}
\newcommand{\Hbeta}{\ifmmode {\rm H}\beta \else H$\beta$\fi}
\newcommand{\Hgamma}{\ifmmode {\rm H}\gamma \else H$\gamma$\fi}
\newcommand{\Hdelta}{\ifmmode {\rm H}\delta \else H$\delta$\fi}
\newcommand{\Lya}{\ifmmode {\rm Ly}\alpha \else Ly$\alpha$\fi}
\newcommand{\Lyb}{\ifmmode {\rm Ly}\beta \else Ly$\beta$\fi}
\newcommand{\HeI}{\ifmmode {\rm He}\,{\sc i}\,\lambda5876 \else 
	          He\,{\sc i}\,$\lambda5876$\fi}
\newcommand{\HeII}{\ifmmode {\rm He}\,{\sc ii}\,\lambda4686 \else 
	           He\,{\sc ii}\,$\lambda4686$\fi}

\newcommand{\mgii}{Mg\,{\sc ii}}

\newcommand{\ciii}{\ifmmode {\rm C}\,{\sc iii} \else C\,{\sc iii}\fi}
\newcommand{\civ}{C\,{\sc iv}}

\newcommand{\mbh}{$M_{\rm BH}$\ }

\newcommand{\msigma}{$M_{\rm BH}$--$\sigma_{*}$\ }

\begin{document}

\title{The Sloan Digital Sky Survey Reverberation Mapping Project: \Halpha \ and \Hbeta \ Reverberation Measurements from first-year spectroscopy and photometry}

\author{C.~J.~Grier\altaffilmark{1,2},
J.~R.~Trump\altaffilmark{1,3},
Yue~Shen\altaffilmark{4,5,*}, 
Keith~Horne\altaffilmark{6}, 
Karen~Kinemuchi\altaffilmark{7},
Ian~D.~McGreer\altaffilmark{8},
D.~A.~Starkey\altaffilmark{6}, 
W.~N. Brandt\altaffilmark{1,2,9}, 
P.~B.~Hall\altaffilmark{10},
C.~S.~Kochanek\altaffilmark{11,12}, 
Yuguang Chen\altaffilmark{13} , 
K.~D.~Denney\altaffilmark{11, 12, 14}, 
Jenny~E.~Greene\altaffilmark{15},  
L.~C.~Ho\altaffilmark{16,17},
Y.~Homayouni\altaffilmark{3},
Jennifer~I-Hsiu~Li\altaffilmark{4}, 
Liuyi~Pei\altaffilmark{4}, 
B.~M.~Peterson\altaffilmark{11,12,18}, 
P.~Petitjean\altaffilmark{19}, 
D.~P.~Schneider\altaffilmark{1,2}, 
Mouyuan~Sun\altaffilmark{20}, 
Yusura~AlSayyad\altaffilmark{15}, 
Dmitry Bizyaev\altaffilmark{7,21}
Jonathan~Brinkmann\altaffilmark{7}, 
Joel~R.~Brownstein\altaffilmark{22},
Kevin~Bundy\altaffilmark{23}, 
K~S.~Dawson\altaffilmark{22}, 
Sarah~Eftekharzadeh\altaffilmark{24}, 
J.~G.~Fernandez-Trincado\altaffilmark{25, 26}, 
Yang~Gao\altaffilmark{27,28}, 
Timothy~A.~Hutchinson\altaffilmark{22}, 
Siyao~Jia\altaffilmark{29}, 
Linhua~Jiang\altaffilmark{16}, 
Daniel Oravetz\altaffilmark{7},
Kaike Pan\altaffilmark{7},
Isabelle~Paris\altaffilmark{30}, 
Kara~A.~Ponder\altaffilmark{31}, 
Christina~Peters\altaffilmark{32},  
Jesse~Rogerson\altaffilmark{33}, 
Audrey Simmons\altaffilmark{7}, 
Robyn~Smith\altaffilmark{34}, 
and Ran~Wang\altaffilmark{16}
}

\altaffiltext{1}{Dept. of Astronomy and Astrophysics, The Pennsylvania State University, 525 Davey Laboratory, University Park, PA 16802}
\altaffiltext{2}{Institute for Gravitation and the Cosmos, The Pennsylvania State University, University Park, PA 16802, USA} 
\altaffiltext{3}{Department of Physics, University of Connecticut, 2152 Hillside Rd Unit 3046, Storrs, CT 06269, USA} 
\altaffiltext{4}{Department of Astronomy, University of Illinois at Urbana-Champaign, Urbana, IL 61801, USA} 
\altaffiltext{5}{National Center for Supercomputing Applications, University of Illinois at Urbana-Champaign, Urbana, IL 61801, USA} 
\altaffiltext{6}{SUPA Physics and Astronomy, University of St. Andrews, Fife, KY16 9SS, Scotland, UK} 
\altaffiltext{7}{Apache Point Observatory and New Mexico State University, P.O. Box 59, Sunspot, NM, 88349-0059, USA} 
\altaffiltext{8}{Steward Observatory, The University of Arizona, 933 North Cherry Avenue, Tucson, AZ 85721, USA} 
\altaffiltext{9}{Department of Physics, 104 Davey Lab, The Pennsylvania State University, University Park, PA 16802, USA} 
\altaffiltext{10}{Department of Physics and Astronomy, York University, Toronto, ON M3J 1P3, Canada}  
\altaffiltext{11}{Department of Astronomy, The Ohio State University, 140 W 18th Avenue, Columbus, OH 43210, USA} 
\altaffiltext{12}{Center for Cosmology and AstroParticle Physics, The Ohio State University, 191 West Woodruff Avenue, Columbus, OH 43210, USA} 
\altaffiltext{13}{Cahill Center for Astronomy and Astrophysics, California Institute of Technology, 1200 E California Blvd., MC 249-17, CA 91125, USA} 
\altaffiltext{14}{Illumination Works, LLC, 5550 Blazer Parkway, Dublin, OH, 43017, USA} 
\altaffiltext{15}{Department of Astrophysical Sciences, Princeton University, Princeton, NJ 08544, USA} 
\altaffiltext{16}{Kavli Institute for Astronomy and Astrophysics, Peking University, Beijing 100871, China} 
\altaffiltext{17}{Department of Astronomy, School of Physics, Peking University, Beijing 100871, China} 
\altaffiltext{18}{Space Telescope Science Institute, 3700 San Martin Drive, Baltimore, MD 21218, USA} 
\altaffiltext{19}{Institut d'Astrophysique de Paris, Universit\'{e} Paris 6-CNRS, UMR7095, 98bis Boulevard Arago, 75014 Paris, France}
\altaffiltext{20}{Key Laboratory for Research in Galaxies and Cosmology, Center for Astrophysics, Department of Astronomy, University of Science and Technology of China, Chinese Academy of Sciences, Hefei, Anhui 230026, China} 
\altaffiltext{21}{Sternberg Astronomical Institute, Moscow State University, Moscow, Russia}
\altaffiltext{22}{Department of Physics and Astronomy, University of Utah, 115 S. 1400 E., Salt Lake City, UT 84112, USA} 
\altaffiltext{23}{UCO/Lick Observatory, University of California, Santa Cruz, 1156 High St. Santa Cruz, CA 95064, USA}
\altaffiltext{24}{Department of Physics and Astronomy, University of Wyoming, Laramie, WY 82071, USA} 
\altaffiltext{25}{Departamento de Astronom\'ia, Casilla 160-C, Universidad de Concepci\'on, Concepci\'on, Chile} 
\altaffiltext{26}{Institut Utinam, CNRS UMR6213, Univ. Bourgogne Franche-Comt\'e, OSU THETA , Observatoire de Besan\c{c}on, BP 1615, 25010 Besan\c{c}on Cedex, France }
\altaffiltext{27}{Department of Engineering Physics and Center for Astrophysics, Tsinghua University, Beijing 100084, China; Key Laboratory of Particle and Radiation Imaging (Tsinghua University), Ministry of Education, Beijing 100084, China} 
\altaffiltext{28}{Department of Engineering Physics and Center for Astrophysics, Tsinghua University, Beijing 100084, China; Key Laboratory of Particle and Radiation Imaging (Tsinghua University), Ministry of Education, Beijing 100084, China} 
\altaffiltext{29}{Department of Astronomy, University of California, Berkeley, CA 94720, USA} 
\altaffiltext{30}{Aix-Marseille Universit\'{e}, CNRS, LAM (Laboratoire d'Astrophysique de Marseille) UMR 7326, 13388, Marseille, France} 
\altaffiltext{31}{Pittsburgh Particle Physics, Astrophysics, and Cosmology Center (PITT PACC), Physics and Astronomy Department, University of Pittsburgh, Pittsburgh, PA 15260, USA} 
\altaffiltext{32}{Dunlap Institute \& Department of Astronomy and Astrophysics, University of Toronto, 50 St George Street, Toronto, ON M5S 3H4 Canada} 
\altaffiltext{33}{Canada Aviation and Space Museum, 11 Aviation Parkway, Ottawa, ON, K1K 4Y5, Canada} 
\altaffiltext{34}{Department of Astronomy, University of Maryland, Stadium Drive, College Park, MD 20742-2421, USA} 
\altaffiltext{*}{Alfred P. Sloan Research Fellow}

\begin{abstract}
We present reverberation mapping results from the first year of combined spectroscopic and photometric observations of the Sloan Digital Sky Survey Reverberation Mapping Project. We successfully recover reverberation time delays between the $g+i$- band emission and the broad \Hbeta\ emission line for a total of 44 quasars, and for the broad \Halpha\ emission line in 18 quasars. { \color{black} Time delays are computed using the {\tt JAVELIN} and {\tt CREAM} software and the traditional interpolated cross-correlation function (ICCF):  Using well-defined criteria, we report measurements of 32 \Hbeta\ and 13 \Halpha\ lags with {\tt JAVELIN},  42 \Hbeta\ and 17 \Halpha\ lags with {\tt CREAM}, and 16 \Hbeta\ and 8 \Halpha\ lags with the ICCF. Lag values are generally consistent among the three methods, though we typically measure smaller uncertainties with {\tt JAVELIN} and {\tt CREAM} than with the ICCF, given the more physically motivated light curve interpolation and more robust statistical modeling of the former two methods.} The median redshift of our \Hbeta-detected sample of quasars is 0.53, significantly higher than that of the previous reverberation mapping sample. We find that in most objects, the time delay of the \Halpha\ emission is consistent with or slightly longer than that of \Hbeta. We measure black hole masses using our measured time delays and line widths for these quasars. These black hole mass measurements are { \color{black} mostly} consistent with expectations based on the local \msigma\ relationship, and are also consistent with single-epoch black hole mass measurements. This work increases the current sample size of reverberation-mapped active galaxies by about two-thirds and represents the first large sample of reverberation mapping observations beyond the local universe ($z < 0.3$).
\end{abstract}


\section{INTRODUCTION}
\label{introduction}
Over the past few decades, the technique of reverberation mapping (RM; e.g., \citealt{Blandford82}; \citealt{Peterson04}) has emerged as a powerful tool for measuring black hole masses ($M_{\rm BH}$) in active galactic nuclei (AGN). RM allows a measurement of the size of the broad line-emitting region (BLR), which is photoionized by continuum emission from closer to the black hole (BH). Variability of the continuum is echoed by the BLR after a time delay that corresponds to the light travel time between the continuum-emitting region and the BLR; this time delay provides a measurement of the distance between the two regions and thus a characteristic size for the BLR ($R_{\rm BLR}$). 

Assuming that the motion of the BLR gas is dominated by the gravitational field of the central BH, we can combine $R_{\rm BLR}$ with the broad emission-line width ($\Delta V$) to measure a BH mass of 
\begin{equation}
M_{\rm BH} =  \frac{fR_{\rm BLR}\Delta V^2}{G},
\label{eq:eq1}
\end{equation}
where the dimensionless scale factor $f$ accounts for the orientation, kinematics, and structure of the BLR. 
  
Thus far, about 60 AGN have \mbh measurements obtained through reverberation mapping (e.g., \citealt{Peterson04}; \citealt{Kaspi00, Kaspi05}; \citealt{Bentz09c, Bentz10a}; \citealt{Denney10}; \citealt{Barth15}; \citealt{Grier12b}; \citealt{Du14, Du16a, Du16b}; \citealt{Hu15}). \cite{Bentz15} provide a running compilation of these measurements\footnote{\texttt{http://www.astro.gsu.edu/AGNmass/}}.  Due to the stringent observational requirements of RM measurements, the existing sample is mainly composed of nearby ($z < 0.3$), lower-luminosity, nearby AGN that have sufficiently short time delays to be measurable with a few months of monitoring using a modest-sized telescope. Because they are low-redshift, these studies typically focus on the \Hbeta \ emission line and other nearby lines in the observed-frame optical. 

RM measurements have established the radius-luminosity ($R-L$) relationship (e.g., \citealt{Kaspi07}; \citealt{Bentz13}), which allows one to estimate the BLR size with a single spectrum and thus estimate \mbh for large numbers of quasars at greater distances where traditional RM campaigns are impractical (e.g., \citealt{Shen11}). However, the current RM sample may be biased; beyond the fact that these AGN are low-redshift, they do not span the full range of AGN emission-line properties (see Figure 1 of \citealt{Shen15}). In addition, the $R-L$ relation is only well-calibrated for \Hbeta, but most higher redshift, single-epoch \mbh estimates are made using \civ \ or \mgii. There are only a handful of RM measurements for \civ, particularly at high redshift (e.g., \citealt{Kaspi07}), and only a few reliable \mgii \ lag measurements have been reported (\citealt{Metzroth06}; \citealt{Shen16}). Such measurements are difficult to make, as higher-luminosity quasars have longer time delays and larger time dilation factors and thus require observations spanning years rather than months. 

The Sloan Digital Sky Survey Reverberation Mapping Project (SDSS-RM) is a dedicated multi-object RM program that began in 2014 (see \citealt{Shen15} for details). The major goals of this program are to expand the number of reverberation-mapped AGN, the range of AGN parameters spanned by the RM sample, the redshift and luminosity range of the RM sample, and to firmly establish $R-L$ relationships for \civ \ and \mgii. SDSS-RM started as an ancillary program of the SDSS-III survey (\citealt{Eisenstein11}) on the SDSS 2.5-m telescope (\citealt{Gunn06}), monitoring 849 quasars in a single field with the Baryon Oscillation Spectroscopic Survey (BOSS) spectrograph (\citealt{Dawson13}; \citealt{Smee13}). Additional photometric data were acquired with the 3.6-m Canada-France-Hawaii Telescope (CFHT) and the Steward Observatory 2.3-m Bok telescope to improve the cadence of the continuum light curves. Observations for the program have continued in 2015, 2016, and 2017 as part of SDSS-IV (\citealt{Blanton17}) to extend the temporal baseline of the program. 

While the primary goals of this program are to obtain RM measurements for $\gtrsim$100 quasars, we have been pursuing a wide variety of ancillary science goals as well, ranging from studies of emission-line and host-galaxy properties to the variability of broad absorption lines (\citealt{Grier15}; \citealt{Shen15b, Shen16b}; \citealt{Sun15}; \citealt{Matsuoka15}; \citealt{Denney16a}). The first RM results from this program were reported by \cite{Shen16}, who measured emission-line lags in the \Hbeta \ and \mgii \ emission lines in 15 of the brightest, relatively low-redshift sources in our sample using the first year of SDSS-RM spectroscopy alone (i.e., no photometric data were used). { \color{black} \cite{Li17a} also measured composite RM lags using a low-luminosity subset and the first year of spectroscopy.} 

We here report results based on the combined spectroscopic and imaging data from the first year of observations, focusing on the \Hbeta \ and \Halpha \ emission lines in the low-redshift ($z < 1.1$) subset of the SDSS-RM sample. We detect significant lags in about {\color{black} 20\%} of our sample. In Section~\ref{sec:data}, we describe the sample of quasars in our study, present details of the data, and discuss data preparation. We discuss our time-series analysis methods in Section~\ref{sec:timeseries}, our results in Section~\ref{sec:discussion}, and summarize our findings in Section~\ref{sec:summary}. Throughout this work, we adopt a $\Lambda$CDM cosmology with $\Omega_{\Lambda}$~=~0.7, $\Omega_{M}$~=~0.3, and $h$~=~0.7. 

\section{DATA AND DATA PROCESSING} 
\label{sec:data}

\subsection{The Quasar Sample} 
\label{sec:sample}
\begin{figure}
\begin{center}
\includegraphics[scale = 0.365, angle = 0, trim = 0 0 0 0, clip]{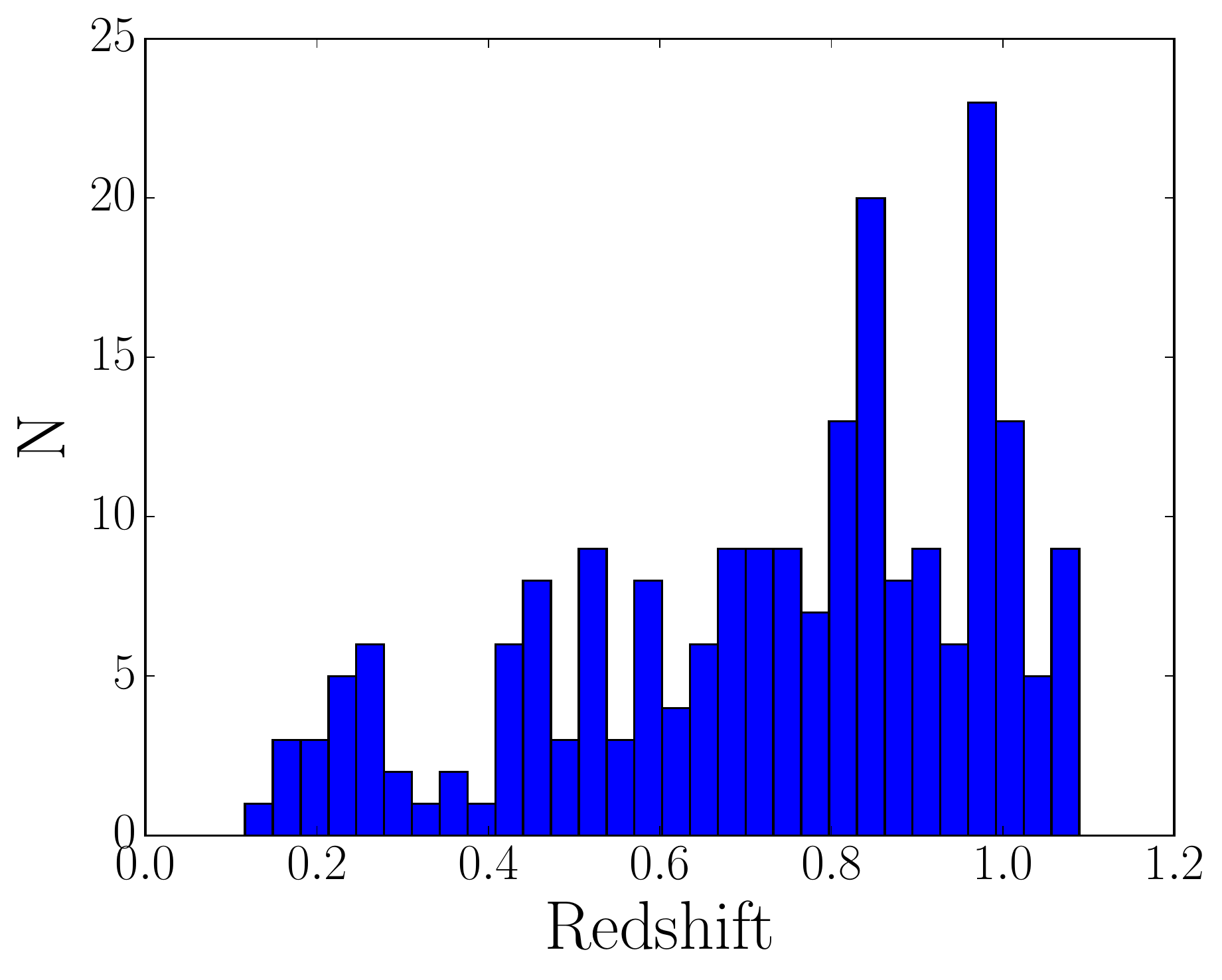}
\includegraphics[scale = 0.365, angle = 0, trim = 0 0 0 0, clip]{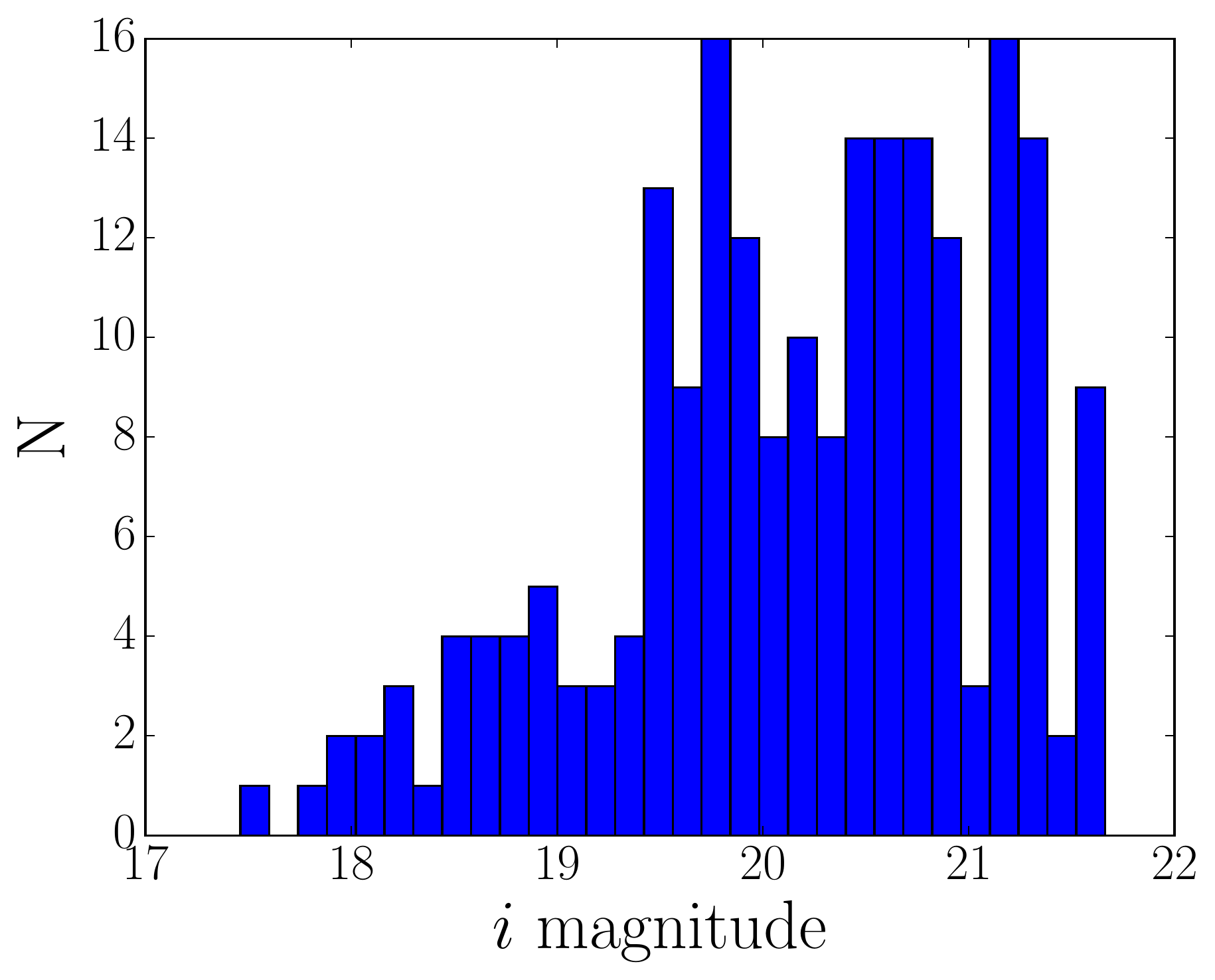}
\includegraphics[scale = 0.365, angle = 0, trim = 0 0 0 0, clip]{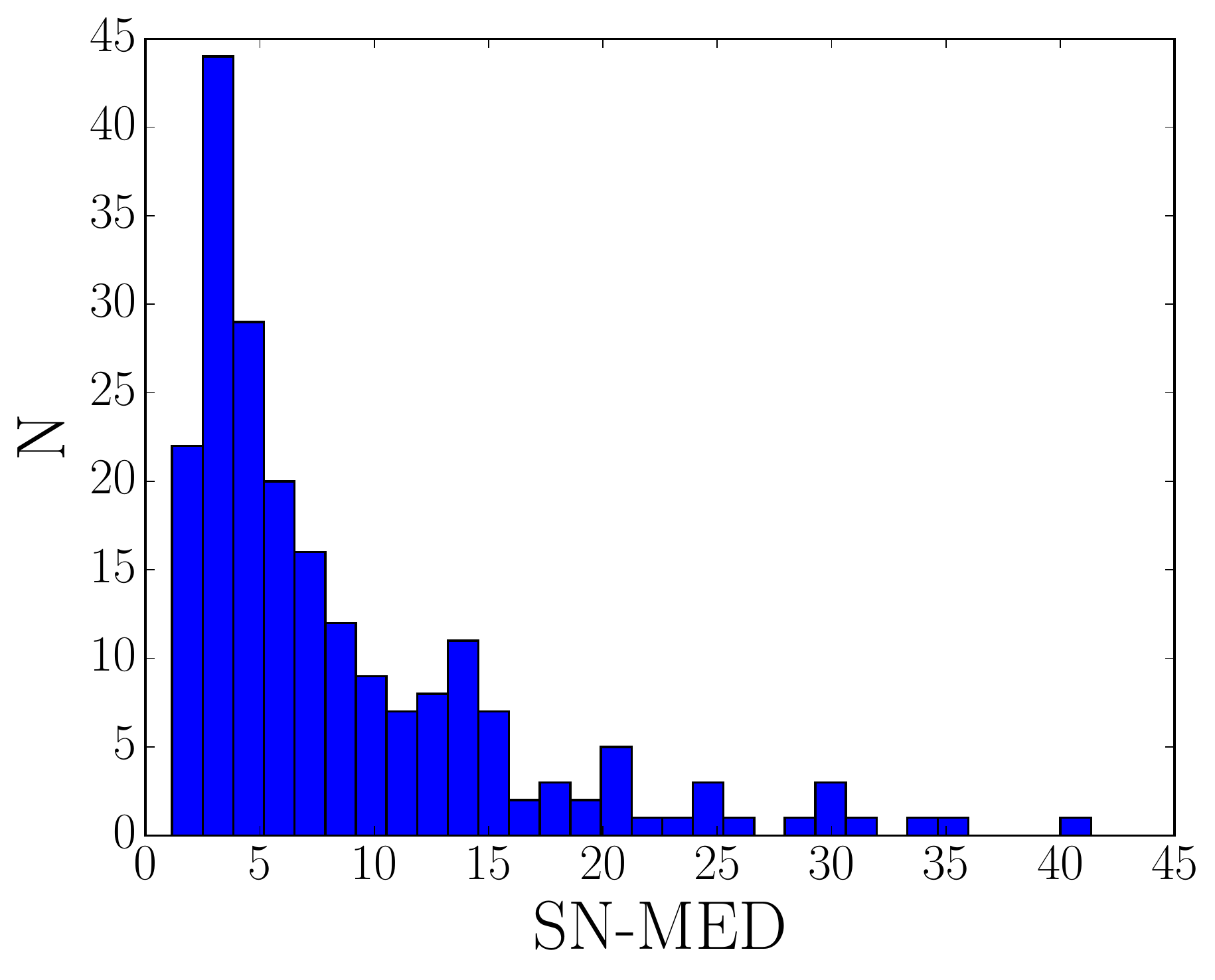}
\includegraphics[scale = 0.365, angle = 0, trim = 0 0 0 0, clip]{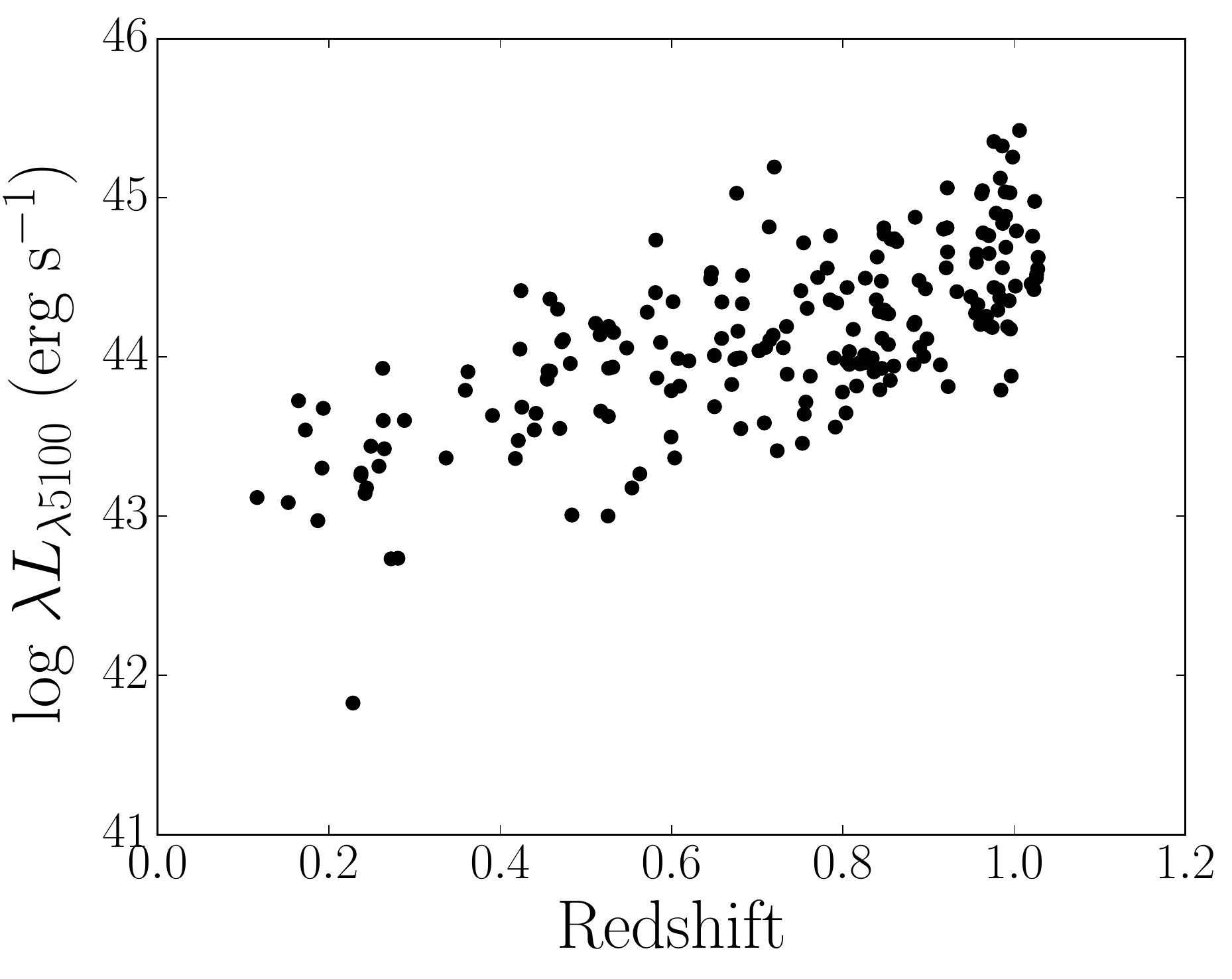}
\caption{From top to bottom: The distributions of our sample of quasars in redshift, $i$ magnitude, median SN-MED (see Section~\ref{sec:sample}), and $\lambda L_{\lambda 5100}$ (the host-subtracted quasar continuum luminosity at 5100\,\AA) as a function of redshift. }
\label{fig:sample}
\end{center}
\end{figure}

We selected our objects from the full SDSS-RM quasar sample, which is flux-limited ($i < 21.7$; measurements by \citealt{Ahn14}) and contains 849 quasars with redshifts of $0.1 < z < 4.5$. A complete description of the parent sample and the properties of the quasars will be reported by \cite{Shen17}.  Within the full sample, there are 222 quasars in the 0.11$~<~z~<~$1.13 redshift range that places \Hbeta \ in the wavelength range of the SDSS spectra. Basic information on these quasars is given in Table~\ref{Table:sample}, including several spectral measurements made by \cite{Shen15b}.  Figure~\ref{fig:sample} presents the distributions of the quasars in redshift, magnitude, typical spectral signal-to-noise ratio (SNR) and luminosity. Of the 222 quasars, 55 are at low enough redshifts ($z < 0.6$) for \Halpha \ to fall within the observed wavelength range of the spectra as well. 

\subsection{Spectroscopic Data}
\label{sec:prepspec} 
\begin{figure}
\begin{center}
\includegraphics[scale = 0.4, angle = 0, trim = 0 0 0 0, clip]{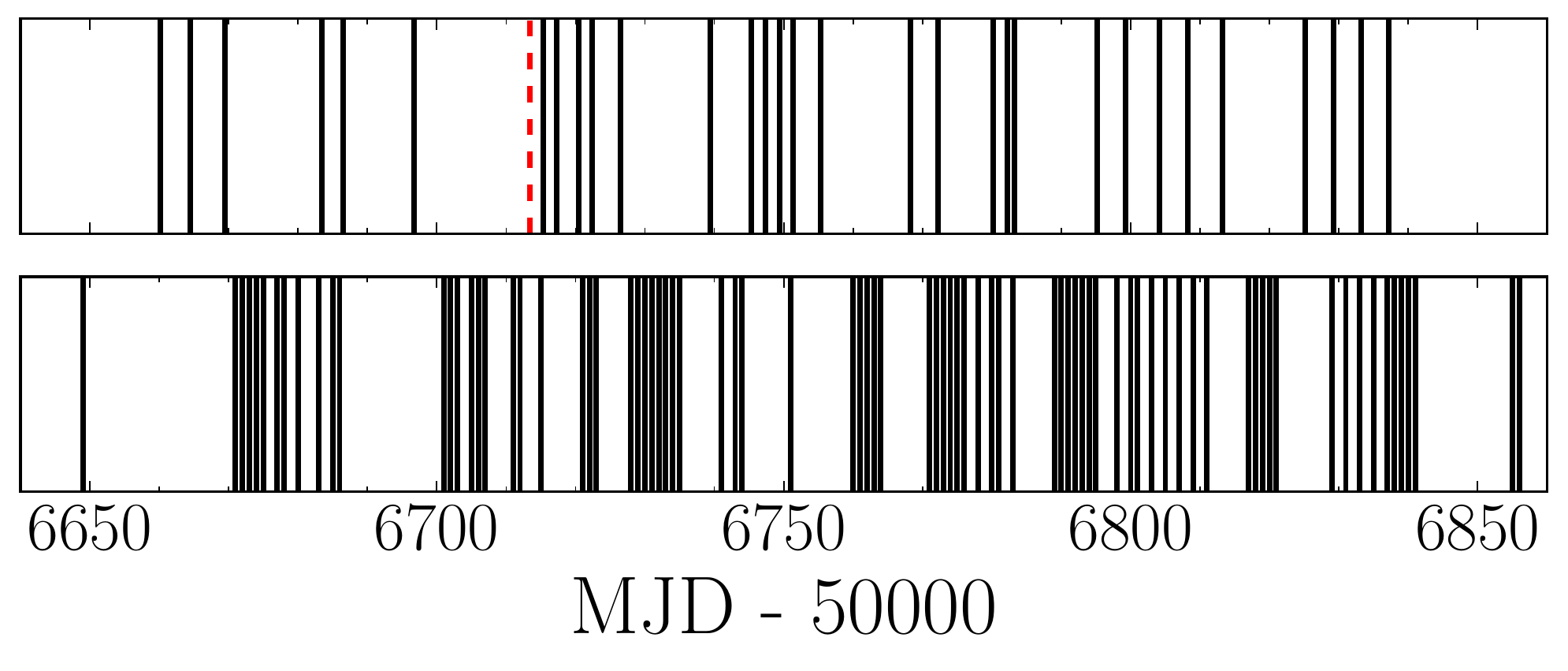}
\caption{The observing cadence for the spectroscopic observations (top panel) and photometric observations (bottom panel). Each vertical black line represents an observed epoch. 
The seventh spectroscopic epoch, shown as a red dashed line, has much lower SNR and is frequently an outlier in the light curves, and so is excluded from our analysis. }
\label{fig:dateplot}
\end{center}
\end{figure}

The SDSS-RM spectroscopic data utilized in this work were all acquired with the BOSS spectrograph between 2014 January and 2014 July. The BOSS spectrograph covers a wavelength range of $\sim 3650-10400$ \AA \ and has a spectral resolution of $R \sim 2000$. The processed spectra are binned to 69 \kms \ pixel$^{-1}$. We obtained a total of 32 spectroscopic epochs with a median of 4.0 days between observations and a maximum separation of 16.6 days. The observations were scheduled during dark time and occasionally had interruptions due to weather or scheduling constraints, so the cadence of the observations varies somewhat throughout the season. Figure~\ref{fig:dateplot} shows the actual observing cadence. The typical exposure time was 2 hours. The data were processed by the SDSS-III pipeline and then further processed using a custom flux-calibration scheme described in detail by \cite{Shen15}. We measure the median SNR per pixel in each epoch for each source, and take the median among all epochs as our measure of the overall SNR for each source, which we designate as SN-MED. The distribution of SN-MED for our sample is shown in Figure~\ref{fig:sample}. 

To improve our relative flux calibrations and produce light curves, we employ a series of custom procedures as implemented in a code called PrepSpec, which is described in detail by \cite{Shen15, Shen16}. A key feature of PrepSpec is the inclusion of a time-dependent flux correction calculated by assuming that there is no intrinsic variability of the narrow emission line fluxes over the course of the RM campaign. PrepSpec minimizes the apparent variability of the narrow lines by fitting a model to the spectra that includes intrinsic variations in both the continuum and broad emission lines. PrepSpec is similar to recent spectral decomposition approaches (e.g., \citealt{Barth15}), but it is optimized to fit all of the spectra of an object simultaneously and includes this flux calibration correction. The PrepSpec model also incorporates components to account for variations in seeing and small wavelength shifts. PrepSpec produces measurements of line fluxes, mean and root-mean-square residual (RMS) line profiles, line widths, and light curves for each of the model components. We note that the PrepSpec RMS line profiles do not include the continuum, and thus differ from commonly-measured RMS line profiles that often still include the continuum (see Section~\ref{sec:mbh} for details). 

We compute $g$- and $i$-band synthetic photometry from each PrepSpec-scaled spectrum by convolving it with the corresponding SDSS filter response curves (\citealt{Fukugita96}; \citealt{Doi10}). We estimate uncertainties in the synthetic photometric fluxes as the quadratic-sum uncertainties resulting from the measurement errors in the spectrum and errors in the flux-correction factor from PrepSpec. We then later merge these light curves with the photometric light curves to improve the cadence of the continuum light curves (see Section~\ref{sec:intercalibration} below). 
We calculate emission-line light curves directly from the PrepSpec fits. 

Of the 32 available epochs, two (the third and seventh epochs) were acquired under poor observing conditions, resulting in spectra with significantly lower SNRs than the other epochs. Upon inspection, the seventh epoch (MJD 56713) appeared as a significant outlier in a large fraction of the light curves (more than 33\% of the \Hbeta \ light curves). We therefore removed Epoch 7 from all of our spectroscopic light curves. There were also occasional cases of ``dropped" epochs and/or loose fibers; these are cases where the fibers were not plugged correctly or the SDSS pipeline failed to extract a spectrum for various reasons. Loose fibers appear as significant low-flux outliers in the light curves, while dropped epochs appear as epochs with zero flux. We excluded all epochs with zero flux and epochs with loose fibers by rejecting points that were offset from the median flux by more than 5 times the normalized median absolute deviation (NMAD; e.g., \citealt{Maronna06}) of the light curve (this threshold was established by visual inspection; see also \citealt{Sun15} for a discussion of dropped fibers). The final emission-line light curves of all 222 quasars are given in Table~\ref{Table:lc_rm005}. We include all spectroscopic epochs in the table and mark those that were excluded from our analysis with a rejection flag (FLAG~=~1).

\subsection{Photometric Data}    
In addition to spectroscopic monitoring with SDSS, we have been observing the SDSS-RM quasars in both the $g$ and $i$ bands with the Steward Observatory Bok 2.3m telescope on Kitt Peak and the 3.6m Canada-France-Hawaii Telescope (CFHT) on Maunakea. Details of the photometric observations and the subsequent data processing will be presented by K. Kinemuchi et~al.~(in preparation). The Bok/90Prime instrument (\citealt{Williams04}) used for our observations has a $\sim1\degree \times1\degree$  field of view using four 4k$\times$4k CCDs each with a plate scale of 0$\farcs$45 pixel$^{-1}$.  Over 60 nights between Jan and June 2014, largely during bright time, we obtained 31 epochs in $g$-band and 27 epochs in $i$. The CFHT MegaCam instrument (\citealt{Aune03}) also has a $\sim1\degree \times1\degree$ field of view, but has a pixel size of 0$\farcs$187. 
Over the 2014 observing period, we obtained 26 epochs in $g$ and 20 in $i$, with a few additional epochs in each band where only some of the fields were observed. 

To produce photometric light curves, we adopt image subtraction as implemented in the software package ISIS (\citealt{Alard98}; \citealt{Alard00}). The basic procedure is to first align the images and create a reference image by combining the best images (seeing, transparency, sky background). ISIS then alters the point-spread function (PSF) of the reference image and scales the target image in overall flux calibration. It then subtracts the two to leave a ``difference" image with the same flux calibration as the reference image, showing the sources that have changed in flux. We then place a PSF-weighted aperture over each source and measure the residual flux in each of the subtracted images to produce light curves. We separately produced reference images and performed the subtraction for each individual telescope, filter, CCD, and field. 

After the image subtraction was complete, we removed bad measurements/outliers from the photometric light curves --- these include points for sources that have fallen off the edge of the detector in certain epochs, saturated sources (either bright quasars themselves or those near a bright star, which show a large dispersion in flux in the differential photometry), and images affected by passing cirrus or other problems that deviate from the median by $>$ 5 times the NMAD of the light curve. 

While the image-subtraction technique allows one to better compensate for changes in seeing and separate seeing-dependent aperture effects from real variability, the ISIS software takes into account only local Poisson error contributions. There are also systematic uncertainties that are not well-captured by these estimates. We follow the procedure outlined by \cite{Hartman04} and \cite{Fausnaugh16a, Fausnaugh17} to apply corrections to the ISIS uncertainties. We extracted light curves for stars of similar magnitude to the quasars, most of which should be non-variable. After eliminating the few variable stars, we determine an error-rescaling factor necessary for each standard star light curve to be consistent with a constant-flux model and plot this factor as a function of magnitude for each CCD/field combination. This provides an estimated error-rescaling factor as a function of magnitude, which we fit as a polynomial and multiply the error estimates by. Scale factors were typically about a factor of two, but range from $\sim$1 for fainter sources to $\sim$10 for the brightest sources. We did not apply scale factors less than 1 (i.e., we did not reduce any uncertainties from their ISIS-reported values). 

\subsection{Light Curve Inter-Calibration} 
\label{sec:intercalibration} 
We have several individual photometric light curves (one for each telescope/field/CCD observation) and a single synthetic photometric light curve (produced from the spectra) in each band for each quasar. For our analysis, it is necessary to place all of the $g$ and $i$-band light curves from all CCDs/telescopes/fields on the same flux scale; this inter-calibration accounts for different detector properties, different telescope throughputs, and other properties specific to the individual telescopes involved. We assume that the time lag between the $g$- and $i$- band is much smaller than we are able to resolve with our data and thus can be treated as zero for intercalibration purposes.  

We performed this inter-calibration using the Continuum REprocessing AGN MCMC ({\tt CREAM}) software recently developed by \cite{Starkey16}. {\tt CREAM} uses Markov Chain Monte Carlo (MCMC) techniques to model the light curves, assuming that the continuum emission is emitted from a central location and is reprocessed by more distant gas (see \citealt{Starkey16} for a thorough discussion of the technique). {\tt CREAM} fits a model driving light curve $X(t)$ to the $g$ and $i$ band light curves $f_\nu (\lambda , t )$ with an accretion disk response function $\psi \left( \tau | \lambda \right)$. The model is  

\begin{equation}
\label{eq_fnucream}
f_{j}(\lambda , t)= \bar{F}_{j} \left( \lambda \right)  + \Delta F_{j}(\lambda)  \int_0^\infty \psi \left( \tau | \lambda \right) X \left( t- \tau \right) d \tau,
\end{equation}

\noindent where each telescope $j$ is assigned an offset $\bar{F}_{j} \left( \lambda \right)$ and flux scaling parameter $\Delta F_{j}(\lambda)$. The offset and scaling parameters control the inter-calibration of the $g$ or $i$ light curves, from multiple telescopes, onto the same scale. 

These parameters are optimized in the MCMC fit, and the rescaled $g$ and $i$ light curves are calculated from the original light curves using 

\begin{equation}
\label{eq_merge}
f_{j,\, \mathrm{new}}(\lambda , t) = \left( f_{j, \, \mathrm{old}} \left( \lambda , t \right) - \bar{F_j} \right)  \frac{ \Delta F_\mathrm{REF} }{ \Delta F_j } + \bar{F}_\mathrm{REF},
\end{equation}

\noindent where the subscript REF indicates the reference telescope/filter combination, and $j$ is calculated for all telescopes at each $g$ or $i$ wavelength. {\tt CREAM} was initially designed to calculate inter-band continuum lags by fitting the accretion-disk response function $\psi \left( \tau | \lambda \right)$. This function is not required in the merging process here --- we are only interested in the inter-calibration parameters $\bar{F}_{j} \left( \lambda \right)$ and $\Delta F_{j}(\lambda)$. We therefore alter {\tt CREAM} such that it has a delta function response at zero lag
$\psi ({\tau} | \lambda) = \delta \left( \tau - 0 \right)$
\noindent for the continuum light curves in each $g$ and $i$ filter. 

{\tt CREAM}'s MCMC algorithm also rescales the nominal error bars using an extra variance, $V_j$, and scale factor parameters, $f_j$, for each telescope (\citealt{Starkey17}). The rescaled error bars are  

\begin{equation}
\label{eq_varer}
\sigma_{\mathrm{ij}} = \sqrt{ \left( f_j \sigma_\mathrm{old, \, ij} \right)^2 + V_j },
\end{equation}

\noindent where $i = 1...N_j$ is an index running over the $N_j$ data points for telescope $j$. The likelihood function $L_j$ penalizes high values of $V_j$ and $f_j$ in the MCMC chain, and is given by
\begin{equation}
\label{eq_BOFvar}
-2 \mathrm{ln} L_j = N_j \mathrm{ln} \left( 2 \pi \right) + \sum_{i=1}^{N_j} [\mathrm{ln} \sigma_\mathrm{ij}^2 + \left( \frac{D_{\mathrm{ij}} - M_{\mathrm{ij}}}{\sigma_{\mathrm{ij}}} \right)^2],  
\end{equation}

\noindent for data $D_{ij}$ and model $M_{ij}$. This approach provides an additional check/correction on the uncertainties for our continuum light curves. 

\begin{figure}
\begin{center}
\includegraphics[scale = 0.255, angle = 0, trim = 0 0 0 0, clip]{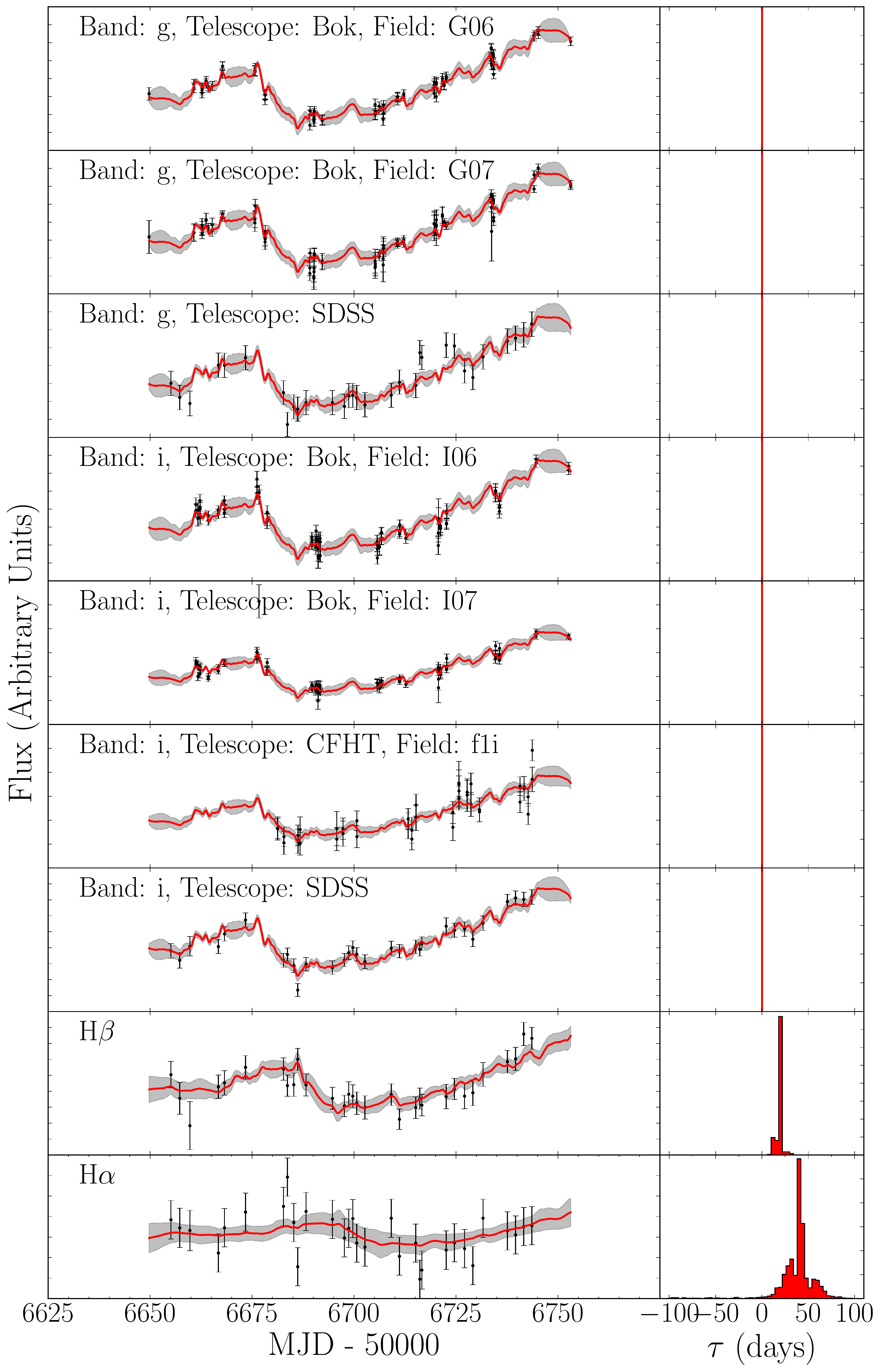}
\caption{{\tt CREAM} model fits to the light curves for SDSS\,J141625.71+535438.5 (RMID 272, $z~=~0.263$) as a demonstration of the inter-calibration technique. Each left panel shows an individual pre-merged light curve (black points) with the {\tt CREAM} model fit and uncertainties in red and gray, respectively. The right panels display the corresponding {\tt CREAM}-calculated posterior distribution of observed-frame time lags calculated for each light curve's response function $\psi(\tau)$. The time lag between the photometric light curves and the synthetic spectroscopic light curves is fixed to zero in order to inter-calibrate the data.}
\label{fig:creamexample}
\end{center} 
\end{figure}

The resulting improved ``merged" light curves from {\tt CREAM} are used in our RM time-series analysis. Figure~\ref{fig:creamexample} presents an example set of light curves for SDSS\,J141625.71+535438.5. The final, intercalibrated light curves for the 222 quasars are provided in Table~\ref{Table:lc_rm005}.  

\section{TIME-SERIES ANALYSIS} 
\label{sec:timeseries} 
\subsection{Lag Measurements} 

Most prior RM measurements have been based upon cross-correlation methods and simple linear interpolation {between observations} (e.g., \citealt{Peterson04}). However, over the past several years, more sophisticated procedures have been developed that model the statistically likely behavior of the light curves in the gaps between observations (e.g., {\tt JAVELIN}, \citealt{Zu11}; and {\tt CREAM}, \citealt{Starkey16}). These procedures provide three key improvements over linear interpolation. Most importantly, their light curves have higher uncertainties in the interpolated regions compared to the observed light curve points, in contrast to the smaller uncertainties between points when using simple linear interpolation.  {\tt JAVELIN} and {\tt CREAM} also use a a damped random walk (DRW) model for the variability, matching observations \citep[e.g.][]{Kelly09,Kozlowski10,MacLeod10}. Finally, they use the same continuum DRW model fit, with a transfer function, to describe the broad-line light curves.  This is essentially a prior that the BLR reverberates (although it allows either a positive or negative reverberation delay).  This assumption is the basic reason that reverberation mapping is possible, although recent observations have also identified periods of non-reverberating variability in NGC~5548 \citep{Goad16}.

We performed our time-series analysis using all three of these methods, with the goal of comparing and contrasting the results from simple interpolation/cross-correlation and different prescriptions for {statistical} modeling of light curves. All of our time-series analysis is performed in the observed frame, and measured time delays are later shifted into the rest frame. { \color{black}  Because our light curves span only about 200 days, we restrict our search to lags from -100 to +100 days. For larger and smaller lags, the overlap between the two light curves is reduced to less than half, making it harder to judge the validity of identifying correlated features. Future data spanning multiple years will soon be able to provide more reliable estimates for longer lags.}

{ \color{black} The most common methods to measure RM time lags are the interpolated cross-correlation function (ICCF; e.g., \citealt{Gaskell87}; \citealt{Peterson04}), and the discrete correlation function (DCF; \citealt{Edelson88}) or $z$-transformed DCF (zDCF; \citealt{Alexander97}). The DCF has been shown to perform best when large numbers of points are present; for cases with lower sampling such as our data, it is better to use the ICCF (\citealt{White94}). The zDCF was designed to mitigate some of the issues with the DCF; however, for this study we opted to use the ICCF, as it is more traditionally used and a detailed comparison between the ICCF and zDCF is not yet available in the literature.} The ICCF method works as follows: For a given time delay $\tau$, we shift the time coordinates of the first light curve by $\tau$ and then linearly interpolate the second light curve to the new time coordinates, measuring the cross-correlation Pearson coefficient $r$ between the two light curves using overlapping points. We next shift the second light curve by $-\tau$ and interpolate the first light curve, and average the two values of $r$. This process is repeated over the entire range of allowed $\tau$, evaluating $r$ at discrete steps in $\tau$. This procedure allows the measurement of $r$ as a function of $\tau$, called the ICCF. The centroid ($\tau_{\rm cent}$) of the ICCF is measured using points surrounding the maximum correlation coefficient $r_{\rm max}$ out to $r \geq 0.8r_{\rm max}$, as is standard for ICCF analysis (e.g., \citealt{Peterson04}). 

We calculated ICCFs and $\tau_{\rm cent}$ for our entire sample of quasars using an interpolation grid spacing of 2 days, calculating the ICCF between $-100$ and 100 days. Following \cite{Peterson04}, we estimate the uncertainty in $\tau_{\rm ICCF}$  using Monte Carlo simulations that employ the flux randomization/random subset sampling (FR/RSS) method. Each Monte Carlo realization randomly selects a subset of the data and alters the flux of each point on the light curves by a random Gaussian deviate scaled to the measurement uncertainty of that particular point. We then calculate the ICCF for the altered set of light curves and measure $\tau_{\rm cent}$ and $\tau_{\rm peak}$. This procedure is repeated 5000 times to obtain the cross correlation centroid distribution (CCCD), and the uncertainties are determined from this distribution. We adopt the median of the distribution as the best $\tau_{\rm ICCF}$ measurement after some modifications and the removal of aliases (described below in Section~\ref{sec:aliases}). { \color{black} Many previous studies adopted the centroid as measured from the actual ICCF rather than the median from the CCCD. However, we use the median of the CCCD because in the case of light curves with lower time sampling, the ICCF centroid can often be an outlier in the CCCD, suggesting that the median of the CCCD is a better characterization of the true lag. However, we do note that for our data, results using the centroid of the ICCF are nearly identical to measurements using the median of the CCCD.}

We used the modeling code {\tt JAVELIN} (\citealt{Zu11, Zu13}) as our primary time-series analysis method. Rather than linearly interpolating between light-curve points, {\tt JAVELIN} models the light curves as an autoregressive process using a damped random walk (DRW) model, and treats the emission-line light curves as scaled, shifted, and smoothed versions of the continuum light curves. The DRW model is observed to be a good description of quasar variability within the time regime relevant to our study (e.g., \citealt{Kelly09}; \citealt{Kozlowski10, Kozlowski16}; \citealt{MacLeod10, MacLeod12}), and so is an effective prior to describe the light curve between observations. {\tt JAVELIN} builds a model of both light curves and simultaneously fits a transfer function, maximizing the likelihood of the model and computing uncertainties using the (Bayesian) Markov chain Monte Carlo (MCMC) technique. The advantage of a method such as {\tt JAVELIN} over the ICCF is that it replaces linear interpolation with a statistically and observationally motivated model of how to interpolate in time. The {\tt JAVELIN} lag measurement takes into account the (increased) uncertainty associated with the interpolation between data points while including the statistically likely behavior of the intrinsic light curve. When multiple light curves of different emission lines are available, {\tt JAVELIN} can model them simultaneously, which improves its performance and helps to eliminate multiple solutions.

The time span of our campaign observations ($\sim$190 days) is shorter than the typical damping timescale of a quasar {($\sim$200-1000~days; \citealt{Kelly09}, \citealt{MacLeod12}, \citealt{Sun15})}, and so {\tt JAVELIN} is unable to constrain this quantity with our data (e.g., \citealt{Kozlowski17}). We thus fix the {\tt JAVELIN} DRW damping timescale to be 300 days (the exact choice of timescale does not matter as long as it is longer than the baseline of our data). We use a top-hat transfer function that is parameterized by a scaling factor, width, and time delay (which we denote as $\tau_{\rm JAV}$) with the width fixed to 2.0 days and the time delay restricted to be within $-100$ to 100 days. The best-fit lag and its uncertainties are calculated from the posterior lag distribution from the MCMC chain.

As discussed in Section~\ref{sec:intercalibration}, \cite{Starkey16} recently developed an alternate approach to modeling light curves and measuring time delays called {\tt CREAM}. In addition to merging the $g$ and $i$ light curves, {\tt CREAM} is also able to infer simultaneously the \Halpha \ and \Hbeta \ lags. To achieve this, we assign a delta function response to the \Halpha \ and \Hbeta \  lags such that
$\psi ({\tau} | \lambda) = \delta \left( \tau - \tau_\mathrm{BLR} \right)$,
\noindent where $\tau_\mathrm{BLR}$ is a fitted parameter in the MCMC chain along with the inter-calibration parameters $\bar{F}_{j} \left( \lambda \right)$ and $\Delta F_{j}(\lambda)$ (see Equation~\ref{eq_fnucream}). {\tt CREAM} self-consistently accounts for the joint errors in calibration and merging of the light curves when determining the lag. The {\tt CREAM} posterior probability histograms for the $\tau_\mathrm{BLR}$ parameters are shown for an example source in Figure~\ref{fig:creamexample}. We again measure the best-fit lag (here denoted $\tau_{\rm CREAM}$) from the posterior lag distribution for the corresponding emission line. 

{ \color{black}  All RM methods operate under the assumption that the broad-line region responds to a ``driving" continuum light curve --- this assumption is generally well-justified given that most monitored AGNs have been observed to reverberate. However, there is a question as to whether or not the 5100\,\AA \ continuum emission is a good proxy for the actual emission driving the emission-line response. We discuss this possible issue in Section~\ref{sec:systematics}. }

\subsection{Alias Identification and Removal} 
\label{sec:aliases} 
Examinations of the CCCD or posterior lag distributions from {\tt JAVELIN} or {\tt CREAM} frequently reveal a clear high-significance peak in the distribution accompanied by additional lower-significance peaks. In general, the presence of multiple peaks or a broad distribution of lags can indicate that the lag is not well-constrained. In some cases, however, one peak is {clearly strongest} and the additional {weaker} peaks are simply aliases resulting from the limited cadence and duration of the light curves. Aliases can sometimes be comparable in strength to the correct time lag, and they often appear in light curves with multiple peaks or troughs. These aliases can skew the $\tau$ measurements and/or produce uncertainties that are extremely large. It is therefore necessary to identify and remove aliases and/or additional secondary peaks to obtain the best lag measurement and associated uncertainty.

\begin{figure}
\begin{center}
\includegraphics[scale = 0.22, angle = 0, trim = 0 0 0 0, clip]{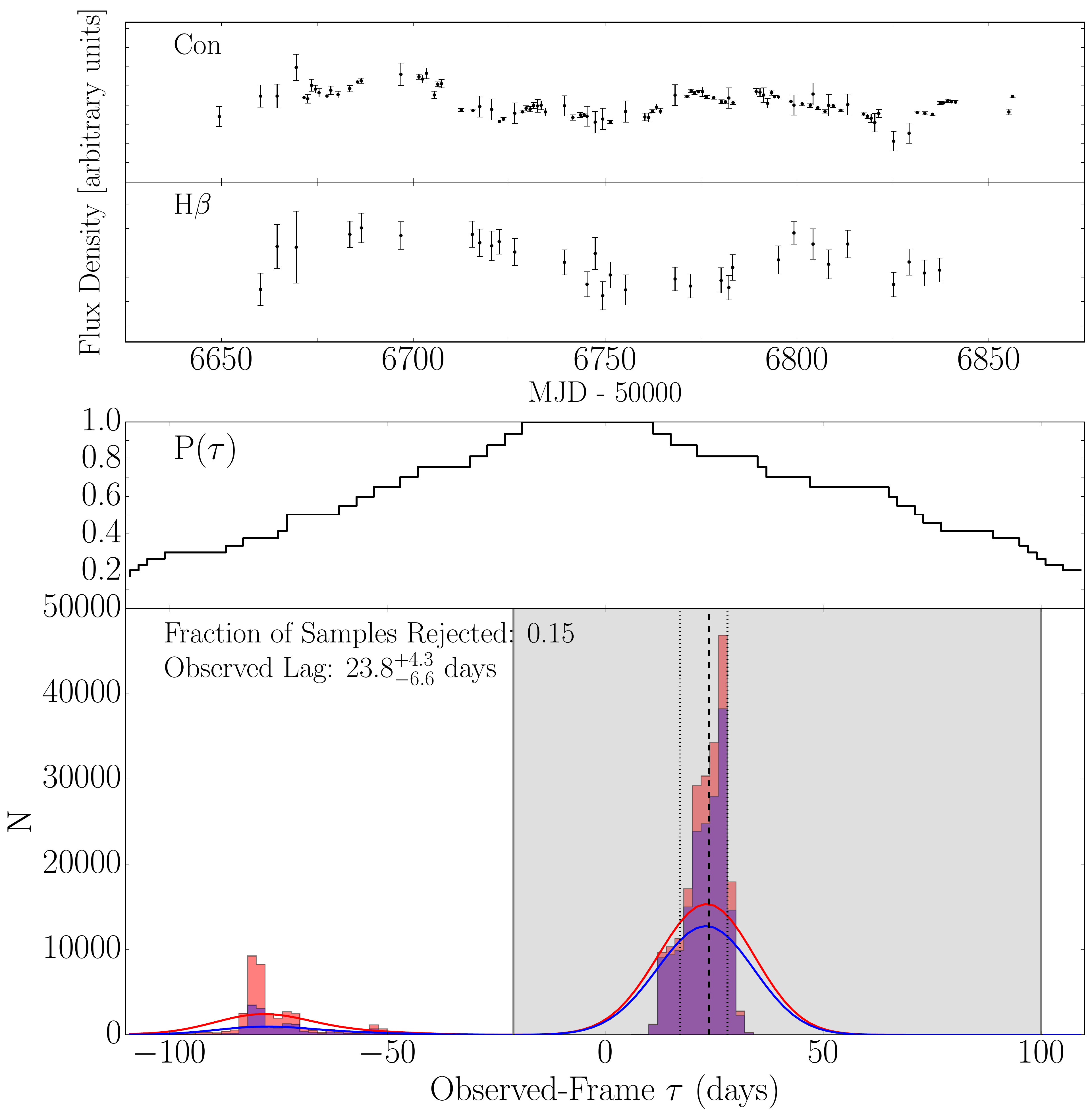} 
\caption{Light curves and the {\tt JAVELIN} posterior \Hbeta \ lag distribution for SDSS\,J141018.04+532937.5 (RMID 229, $z~=~0.470$). The top two panels show the continuum and \Hbeta \ light curve. For display purposes, multiple observations within a single night are averaged and shown as a single point. The third panel from the top shows $P(\tau$) used to weight the posterior lag distribution. The pink shaded histogram shows the {\tt JAVELIN} posterior lag distribution before applying the weights and the purple shaded histogram is the posterior weighted by $P(\tau$); see Section \ref{sec:aliases}. The solid red and blue lines are the smoothed posterior distributions for the unweighted and weighted distributions, respectively. The gray shaded region shows the lag samples surrounding the main peak of the model distribution that were included in the final lag measurement for this source. Vertical black dashed and dotted lines indicate the measured time delay and its uncertainties, respectively, estimated from median and the mean absolute deviation of the lag distribution within the shaded region.}
\label{fig:gmm_example} 
\end{center}
\end{figure}

{Multiple CCCD peaks have been a common feature of previous RM observations, but alias removal in these single-object campaigns was typically applied by visual inspection in an ad-hoc way (B. Peterson, private communication). We instead developed a quantitative technique for alias rejection, appropriate for multi-object RM surveys like SDSS-RM. First,} we applied a weight on the distribution of $\tau$ measurements in the posterior probability distributions that takes into account the number of overlapping spectral epochs at each time delay. If the true lag is so large that shifting by $\tau$ leaves no overlap between the two light curves, then we have a prior expectation that the true lag $\tau$ is not detectable with these data. If shifting one light curve by $\tau$ leaves $N(\tau$) data points in the overlap region, we may expect to be able to detect $\tau$ with a prior probability that is an increasing function of $N(\tau$). We define this weight $P(\tau$) = $[N(\tau)/N(0)]^2$, where $N(0)$ is the number of overlapping points at a time delay of zero. The weight on each $\tau$ measurement is thus 1 for $\tau$ = 0 and decreases each time a data point moves outside the data overlap region when the light curve is shifted, eventually reaching 0 when there is no overlap. {Lags with few overlapping points are less likely to be reliable, since at fixed correlation coefficient $r$ a smaller number of points leads to a higher null-probability $p$.  In this way the $N(\tau)$ prior acts as a conservative check on longer lags, requiring stronger evidence to conclude detections with less light curve overlap.  We tested different exponents for $P(\tau) = ( N(\tau)/N(0) )^k$ and ultimately adopted on $k=2$ based on visual inspection of the apparent lags in the light curves.} Figure~\ref{fig:gmm_example} shows an example of the effect that this weighting has on the posterior lag distributions.

To identify peaks and aliases in the posterior distribution, we smoothed the posterior lag distributions (the cross correlation CCCD or the {\tt JAVELIN}/{\tt CREAM} MCMC posterior lag distributions) by a Gaussian kernel with a width of 5 days (the choice of 5 days was determined by visual inspection). The tallest peak of the smoothed distribution was then identified as the primary lag peak. We searched for local minima on either side of this primary peak and rejected all lag samples that fell outside of these local minima. The lag $\tau$ and its uncertainties were then measured as the median and normalized mean absolute deviation of the remaining lag distribution. We performed this alias-removal procedure on the {\tt JAVELIN} and {\tt CREAM} posteriors and the ICCF CCCDs. Figure~\ref{fig:gmm_example} provides a demonstration of this procedure. We note that the weighting discussed above is only used to select primary peaks and their accompanying lag samples (i.e., identify the range of lags to include); we make our lag measurements from the unweighted posteriors that fall within that lag range. 

\subsection{Lag-Significance Criteria} 
\label{sec:significance} 
In many cases, we find no significant correlation between the two light curves or are otherwise unable to obtain a good measurement of $\tau$ (i.e., the lag is formally consistent with zero when the uncertainties are taken into account). In order to consider the lag a ``significant" detection, we require the following: 
\begin{enumerate} 
\item The measured $\tau$ is formally inconsistent with zero to at least 2$\sigma$-significance (i.e., the absolute value of the lag is greater than twice its lower-bound uncertainty for positive lags and twice its upper-bound uncertainty for negative lags). 
\item Less than half of the samples have been rejected during the alias-identification steps described above; if this alias-removal system excludes more than half of the samples, this is an indication that we lack a solid measurement of $\tau$. 
\item  The maximum ICCF correlation coefficient, $r_{\rm max}$, must be greater than 0.45. This assures that the behavior in the two light curves is well-correlated. This number was determined to remove low-quality lag measurements and retain our highest-quality detections, as determined based on visual inspections of the light curves and the posterior distributions (see Section~\ref{sec:quality} for details). 
\item The continuum and line light-curve RMS variability SNR is greater than 7.5 and 0.5, respectively (see below). This constraint excludes lag measurements that are due to spurious correlations between noisy light curves or long, monotonic trends rather than an actual reverberation signal, and effectively requires that there is significant short-term variability in the light curves. 
\end{enumerate} 

This final criterion requires measurements of the continuum and line light-curve variance. To parameterize this, we define the ``light curve SNR" as the intrinsic variance of the light curve about a fitted linear trend, divided by its uncertainty. First, a linear trend is fit to the light curves. Following \cite{Almaini00} and \cite{Sun15}, we measure the intrinsic variance from the observed $g$-band light curves using a maximum-likelihood estimator to account for the measurement uncertainties. The RMS variation that we observe in the light curves, $\sigma_{\rm obs}$, is a combination of the intrinsic variance $\sigma_{\rm int}$ and the measurement error $\sigma_{\rm err}$, such that $\sigma_{\rm obs}^2 = \sigma_{\rm int}^2 + \sigma_{\rm err}^2$. The maximum-likelihood estimator finds the intrinsic variance that maximizes the likelihood of reproducing the observed variance given the time-dependent error. Sources with short-term variability (i.e., variability other than a smooth trend) will show an excess variance about the fitted linear model, and it is only for these sources that reliable lags can be obtained. 

As with our $r_{\rm max}$ threshold, our chosen light curve SNR thresholds were chosen to remove spurious lag measurements while still retaining all of our highest-quality lag detections. We note, however, that the light curve SNR as measured here is a somewhat coarse measure of the light curve quality for the purpose of lag determination, since it is a measure of the average variability over the entire light curve rather than a measure of short-term variations suitable for a lag measurement. This is why we require a line RMS variability of only 0.5, since many 0.5$<$SNR$<$1 light curves still contain significant short-term variations and a reverberation signal that meets our other criteria. Despite this, the light curve SNR remains a useful way to flag spurious correlations between noisy light curves or long, monotonic variability. 

\begin{figure}
\begin{center}
\includegraphics[scale = 0.35, angle = 0, trim = 0 0 0 0, clip]{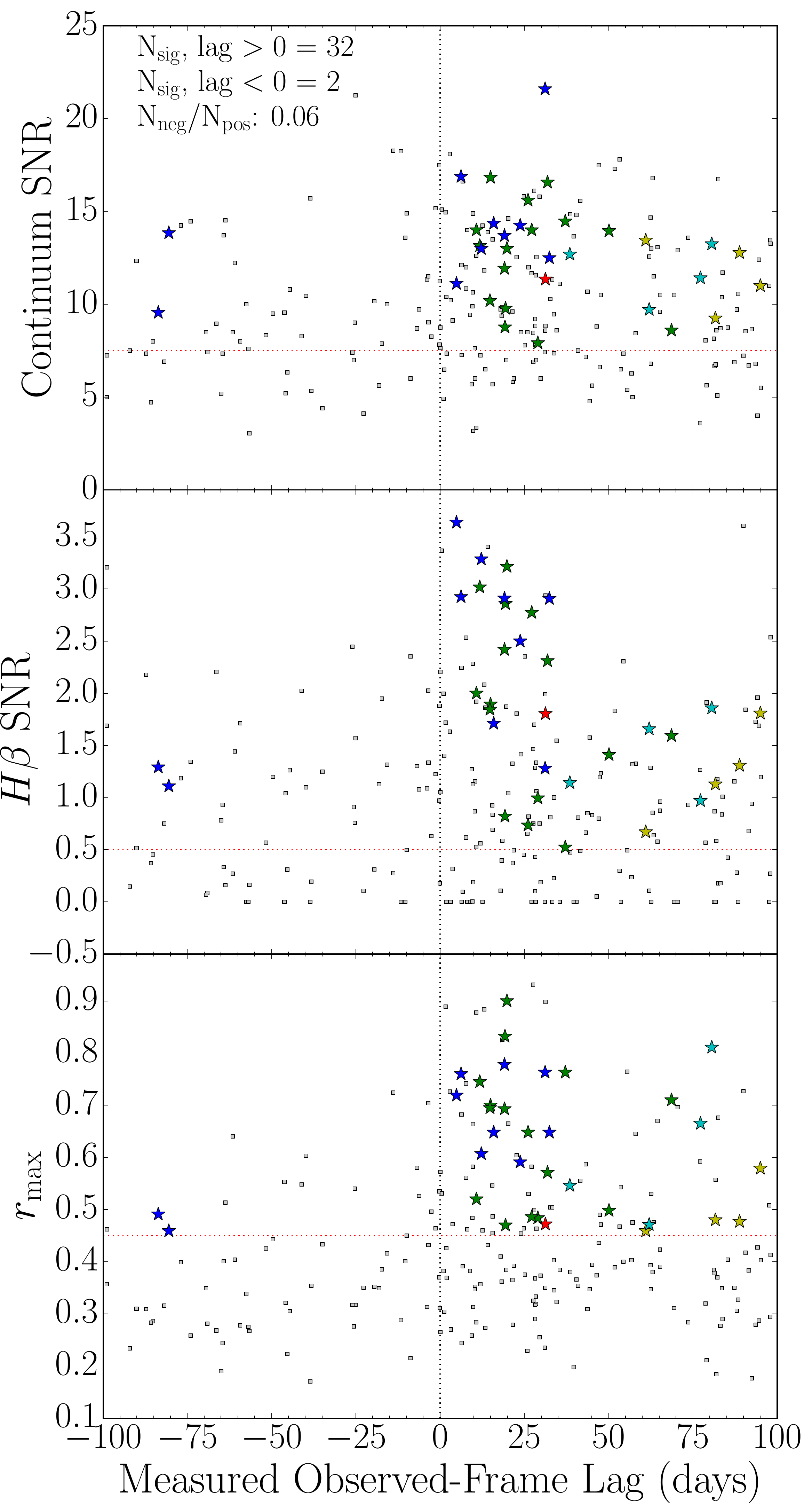}
\caption{Measured time lags vs. various parameters used to determine lag significance for our {\tt JAVELIN} time-series analysis, as discussed in Section~\ref{sec:significance}. The top panel shows the continuum light-curve SNR above a linear trend, the middle panel shows the light-curve variance SNR of the \Hbeta \ light curves, and the bottom panel shows the maximum correlation coefficient of the ICCF, $r_{\rm max}$. Lag measurements that were determined to be significant by our criteria are indicated by stars and are color-coded by the quality rating assigned (see Section~\ref{sec:quality}). Red, yellow, cyan, green, and blue represent measurements with assigned quality ratings of 1, 2, 3, 4, and 5, respectively (red and yellow are the lowest-quality measurements, while blue and green are the highest). The number of significant lags greater than and less than zero is indicated in the figure text. The black vertical dotted line shows a time lag of zero and the red horizontal dotted line shows the cutoff threshold adopted for each parameter.}
\label{fig:falsepositive}
\end{center}
\end{figure}

In order to estimate the false-positive detection rate of each method, we follow \cite{Shen16} and investigate the relative incidence of positive and negative lags. If all lag measurements were due to noise and not due to physical processes, one would expect to find equal numbers of positive and negative lags (we assume that there is no physical reason to measure a negative lag and thus all negative lags are due to the noise and/or sampling properties of our light curves). Figure~\ref{fig:falsepositive} shows the measured \Hbeta \ $\tau_{\rm JAV}$ for all 222 quasars as a function of our various detection threshold parameters. We find that there is a preference for both the detected and non-detected lag measurements to be positive, suggesting that overall, we are measuring more physical lags. We also find that light curves with high intrinsic variability are more likely to show positive lag detections, and there is a strong preference for ``significant" \Hbeta \ lags to be positive, which suggests that, statistically, we are detecting mostly real lag signals. 

Of our significant \Hbeta \ lag detections from {\tt JAVELIN}, {\color{black}  32} are positive and {\color{black}  2} are negative; these negative lags can be considered ``false positives", as they are unphysical from a RM standpoint. Statistically speaking, this suggests that we likely have a similar number of ``false-positive" positive \Hbeta \ lags as well, which is a {\color{black} 6.3$_{ -2.1}^{+7.3}$\%} false-positive rate (calculations of uncertainties follow \citealt{Cameron11}). We thus expect on the order of {\color{black}  30} of our \Hbeta \ lag measurements from {\tt JAVELIN} to be real. We observe a similar fraction of false positives in our \Halpha \ lag measurements (not pictured), with {\color{black}  13} significant positive lags and {\color{black} 1} significant negative lag, corresponding to a false positive rate of {\color{black}  7.7$_{ -2.6}^{+14.0}$\%}. \cite{Shen15} simulated the expected quality of data from the SDSS-RM program (light curve cadence, SNR, etc) and estimated a false-positive rate of between 10--20\%, which is consistent with these estimates. Our criteria for reporting detected lags are quite stringent and are meant to be conservative --- the overall preference for positive lags (both significant and insignificant) suggests that it is likely that we have ``detected" lags in other objects, but the lag measurements themselves were not well-constrained and so they are excluded from our analysis.  

Our false-positive rate is fairly stable to reasonable changes in the parameters used to determine lag significance. Altering the threshold for continuum light-curve SNR (within the ranges of 6--8.5) changes the false-positive rate by less than 3\% (which corresponds to just one additional false-positive measurement), and altering the line light-curve SNR within the ranges of 0.3--0.8 changes the rate by less than a percent. The false-positive rate is more sensitive to $r_{\rm max}$ changes, as varying the $r_{\rm max}$ threshold to values within the range of 0.1--0.5 alters the false-positive rate by 15--20\%. Despite the stability of the false-positive rate, all three criteria place important constraints on the quality of the reported lag measurements, and thus their primary utility is in rejecting poor measurements, both positive and negative. 

\begin{figure*}
\begin{center}
\includegraphics[scale = 0.33, angle = 0, trim = 0 0 0 0, clip]{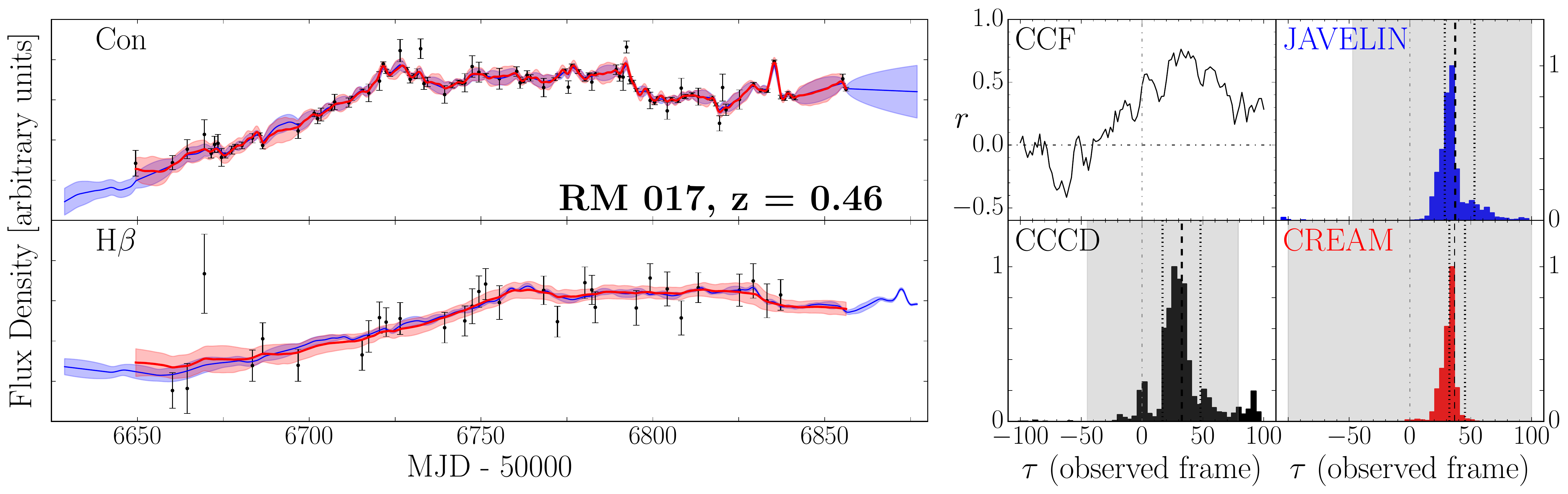}
\caption{Light curves and models for the \Hbeta \ emission line analysis of SDSS\,J141324.28+530527.0 (RMID 017, $z = 0.456$). The continuum and \Hbeta \ light curves are presented in the top and bottom of the left panels. For display purposes, we show the weighted mean of all epochs observed within a single night. The {\tt JAVELIN} model and the uncertainty envelope are given in blue, and the {\tt CREAM} models and their uncertainties in red. The right four panels show the results of the time-series analysis. The top left panel shows the ICCF. The other three panels present the lag distributions for the three different methods, normalized to the tallest peak in the distribution. The bottom left panel shows the CCCD, the top right panel shows the {\tt JAVELIN} posterior lag distribution, and the bottom right panel shows the {\tt CREAM} posterior lag distribution. Black vertical dashed and dotted lines correspond to the measured observed-frame lag and its uncertainties. The gray dash-dotted vertical lines indicate a lag of zero to guide the eye, and the horizontal dash-dotted line in the CCF panel shows a cross-correlation coefficient $r$ of 0. The gray shaded area covers the regions of the posteriors that were included in the measurements, as determined during the alias rejection procedure (see Section~\ref{sec:aliases}). Similar figures for each source are included in the online article.}
\label{fig:hbdetections}
\end{center}
\end{figure*}

\begin{figure*}
\begin{center}
\includegraphics[scale = 0.33, angle = 0, trim = 0 0 0 0, clip]{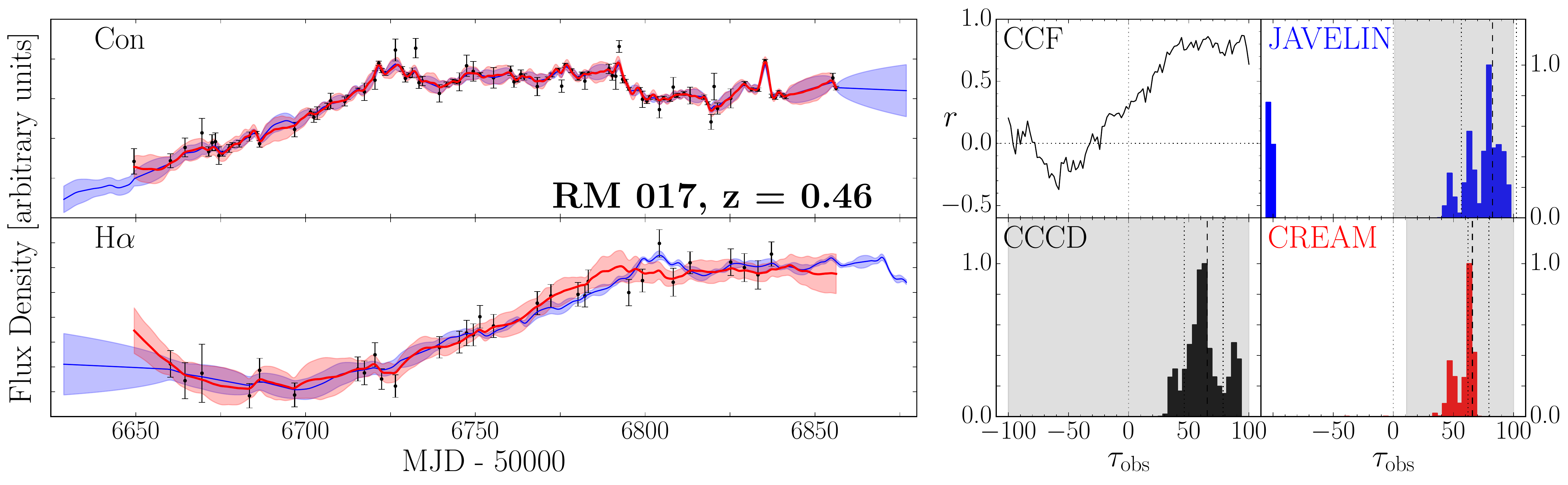}
\caption{Light curves and output for the \Halpha \ time-series analysis for SDSS\,J141324.28+530527.0 (RMID 017, $z = 0.456$). Lines and symbols are the same as in Figure~\ref{fig:hbdetections}. Similar figures for each source are included in the online article.}
\label{fig:hadetections}
\end{center}
\end{figure*}

Having established that the majority of our significant, positive lag detections are likely to be real, we further restrict our significant-lag sample to only those lags that are greater than zero, as a negative lag is unphysical in terms of RM. Our significant lag detections with $\tau~>$~0, detected either by {\tt JAVELIN} or {\tt CREAM}, are reported in Table~\ref{Table:lags}. We also present the light curves and their ICCFs, CCCDs, {\tt JAVELIN} model fits, {\tt JAVELIN} lag posterior distributions, {\tt CREAM} fits, and {\tt CREAM} posterior lag distributions in Figure sets~\ref{fig:hbdetections} and~\ref{fig:hadetections} for all reported positive-lag detections. 

\subsection{Comparison between different lag-detection methods}
\label{sec:methodcomp}
\begin{figure}
\begin{center}
\includegraphics[scale = 0.3, angle = 0, trim = 0 0 0 0, clip]{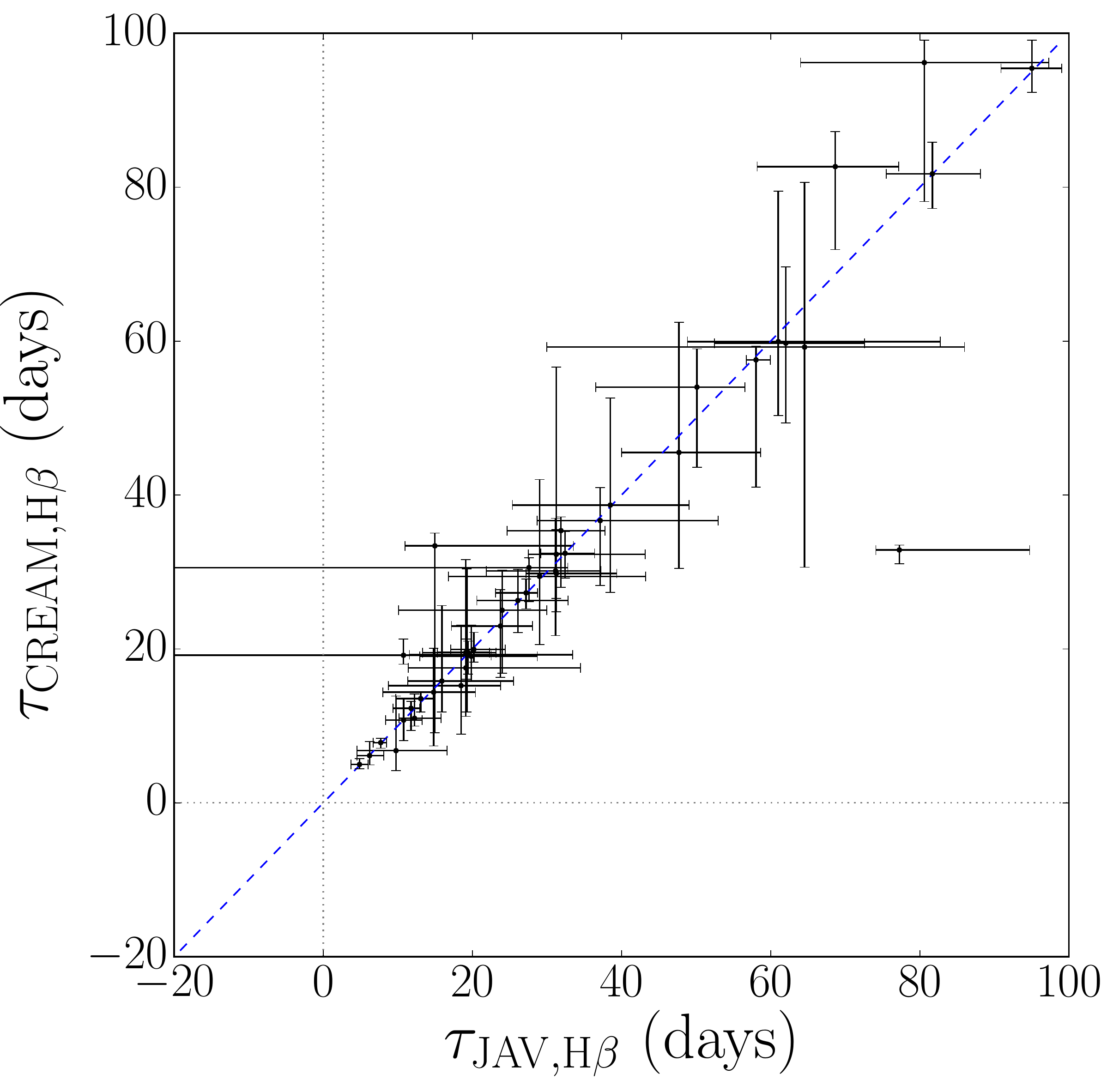}
\includegraphics[scale = 0.3, angle = 0, trim = 0 0 0 0, clip]{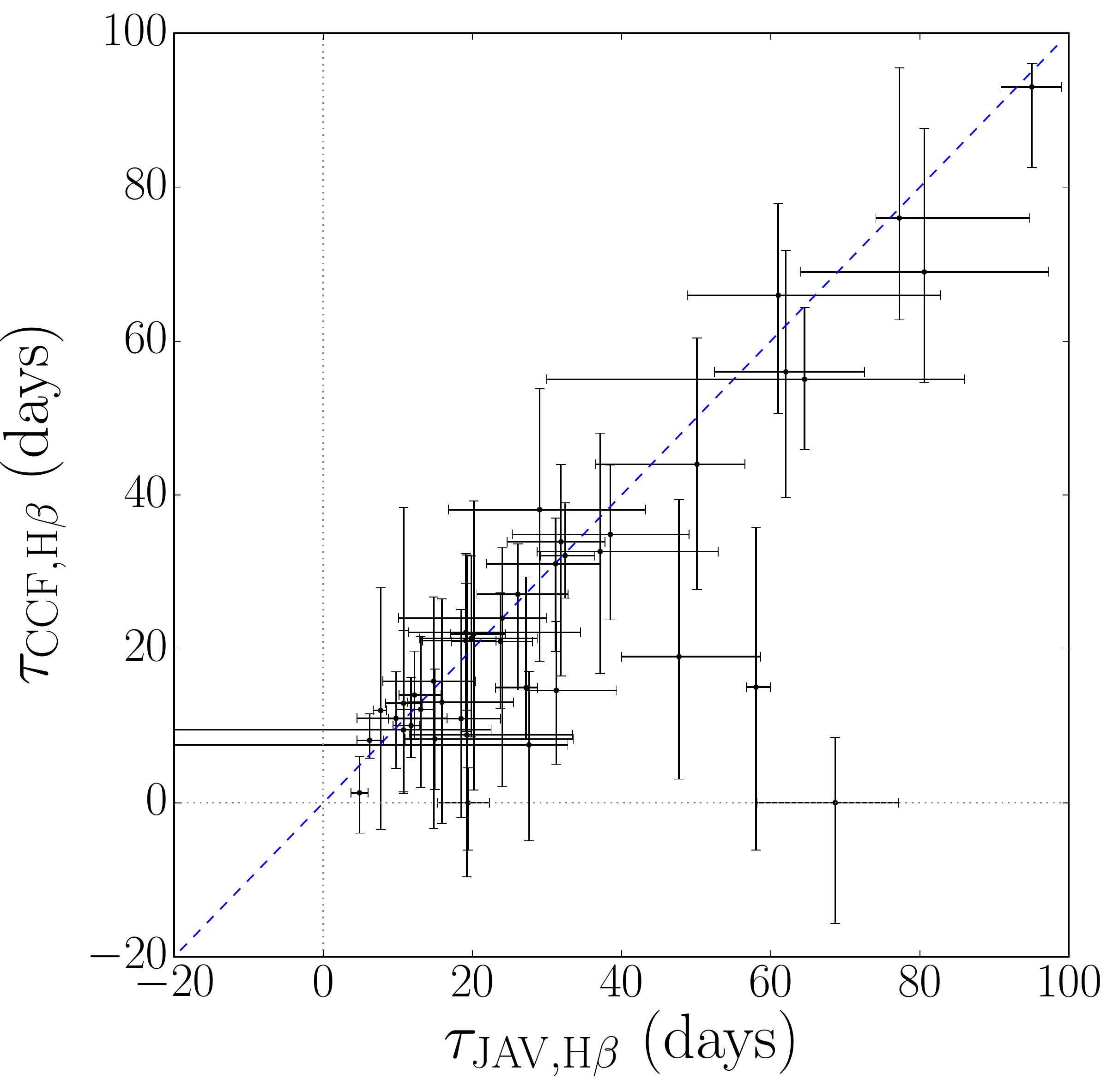}
\caption{Comparison of the observed-frame \Hbeta \ $\tau_{\rm CREAM}$ and $\tau_{\rm JAV}$ measurements (top panel), {and the $\tau_{\rm JAV}$ and $\tau_{\rm ICCF}$ measurements (bottom panel). In both panels} the blue dashed line shows a ratio of one-to-one. { \color{black} Gray dotted lines indicate time lags of zero along both axes to guide the eye.}}
\label{fig:methodcompare}
\end{center}
\end{figure}

One of the aims of our study was to compare results from the three different time-series analysis methods (ICCF, {\tt JAVELIN}, and {\tt CREAM}). The top panel of Figure~\ref{fig:methodcompare} shows that the {\tt JAVELIN} and {\tt CREAM} \Hbeta \ lag measurements are consistent (within 1$\sigma$) for all but one object. Visual inspection of the outlier (RMID 622) indicates that the disagreement can be attributed to the presence of multiple peaks in the posterior distributions. There are peaks in the {\tt JAVELIN} posteriors that match those from {\tt CREAM}, but the peak strength ratios are reversed. 

The agreement with the ICCF results is also generally quite good, as shown by the bottom panel of Figure~\ref{fig:methodcompare}. When the lag is considered detected with the ICCF method, the $\tau_{\rm ICCF}$ measurements are generally consistent with both {\tt JAVELIN} and {\tt CREAM} (i.e., all three methods agree, as these are generally our strongest cases). In the quasars with (poorly detected) ICCF lags that differ from the {\tt JAVELIN} and {\tt CREAM} lags by $>$1$\sigma$, the posteriors of the different methods include the same peaks but at different strengths. The smaller uncertainties and larger number of well-detected lags with {\tt JAVELIN} and {\tt CREAM} is largely due to their use of the same (shifted, scaled, and smoothed) DRW model for both the continuum and broad-line light curves. In contrast, the ICCF assumes independent linearly interpolated light curves for the continuum and broad lines. Well-measured light curves with high sampling result in nearly identical lag measurements from the ICCF and {\tt JAVELIN} (as shown by \citealt{Zu11}), and differences between the methods become apparent only for datasets like SDSS-RM with low cadence and noisy light curves. 

{ \color{black} Inspection of the light curves for quasars with mismatched ICCF lags (e.g., RMID 305 and 309 for \Hbeta, and RMID 779 for \Halpha) show that shifting the emission-line light curves by the {\tt JAVELIN} and {\tt CREAM} lags provides a better match to visual features repeated in both light curves than shifting by the ICCF lags does, and so {\tt JAVELIN} and {\tt CREAM} appear to be more reliable. \cite{Jiang17} have also run simulations with mock light curve data that suggest {\tt JAVELIN} performs better than the ICCF in recovering true lags in the regime of sparsely-sampled light curves. A full simulation comparing the detection completeness/efficiency for BLR lags among these different methods is currently underway (Li et~al., in preparation). However, for our study, the above reasons and visual inspections of the light curves in Figures~6~and~7 support the use of the {\tt JAVELIN} and {\tt CREAM} results for our main lag detections. 
}

Using the same positive/negative lag fraction as a false-positive estimate, we find higher false-positive rates for {\tt CREAM} and the ICCF than we did for {\tt JAVELIN}. For {\tt CREAM}, we measure a false-positive fraction of {\color{black} 16.7$_{-4.2}^{+7.3}$\% for \Hbeta \ (42 positive, 7 negative) \ and 11.8$_{-4.0}^{+12.2}$\% for \Halpha  \ (17 positive, 2 negative). For the ICCF, we measure a fraction of 25$_{ -7.7}^{+13}$\% for \Hbeta \ (16 positive, 4 negative), though we do not measure any significant negative \Halpha \ lags and measure only 8 positive lags, for a false-positive rate of zero (with an upper 1$\sigma$ uncertainty of 18\%). } 

\subsection{Lag-Measurement Quality} 
\label{sec:quality} 
As suggested by our non-zero false-positive rates, it is statistically likely that a few of our lag measurements are false detections. Our objective criteria for significant lag detection minimizes the false-positive rate and removes poor lag measurements, but does not eliminate the possibility for false detections entirely. 

We tested the reliability of our lag estimates with a modified bootstrapping simulation, specifically to test whether or not our lag measurements are strongly dependent on the flux uncertainties of the light curves. For each light curve with $N$ points, we randomly draw epochs $N$ times with replacement, counting how many times each epoch is selected ($n_{\rm select}$). The uncertainty on the flux of each epoch is then multiplied by $1/\sqrt{n_{\rm select}}$ if it is selected at least once --- if the epoch is not selected at all, its uncertainties are doubled. This is done 50 times for each source, creating 50 different iterations of both the continuum and \Hbeta \ light curves. We then run our {\tt JAVELIN} analysis on the light curves with the altered uncertainties and measure the lag. 

From these simulations, we compare the distribution of recovered lags with the original lag measured from the unaltered light curves and determine what percentage are consistent with the original lag to within 1$\sigma$ and 2$\sigma$. We naturally expect 68.3\% of the resampled lags to be consistent to within 1$\sigma$ and 95\% to be consistent to within 2$\sigma$. On average, 81\% of the bootstrap simulations are within 1$\sigma$ of the original lag measurement, and 87\% are within 2$\sigma$. This indicates that the {\tt JAVELIN} lag estimates are robust against the uncertainties in the estimated errors in the light curve fluxes. 

While we have shown that our lag measurements are generally robust, visual inspection leads us still to believe that some lags are more likely to be real than others, so we have assigned quality ratings to each of our lag measurements based on several different factors. The quality ratings range from 1--5, with 1 being the poorest-quality measurements and 5 being the highest-quality detections. When assigning these quality ratings, we paid particular attention to the following: 
\begin{enumerate} 
\item The uni-modality of the posterior distribution: How smooth is this distribution? Are there many other peaks beyond the main peak, or perhaps a lot of low-level noise? 
\item Agreement between different methods: Do all three methods (ICCF, {\tt CREAM}, and {\tt JAVELIN}) result in consistent lags? { \color{black} In about two thirds of our detections, our procedure yielded detected lags using {\tt JAVELIN} or {\tt CREAM} but not using the ICCF. Our statistical analysis (e.g., Figure~\ref{fig:falsepositive}) indicates these lags are real in the statistical sense. The ICCF is likely less powerful in detecting lags in cases where we have lower-SNR and/or lower-cadence light curves, so generally we prefer agreement between {\tt CREAM} and {\tt JAVELIN} only. However, if the ICCF results are also consistent, this likely indicates a more solid measurement, so we take this into account when evaluating the quality of these measurements.}
\item Light-curve variability: Are there apparent short-term variability features in the continuum light curve that are also apparent in the emission-line light curve? Can we identify the lag by eye? Does the reported lag look reasonable if we shift the emission-line light curve by this lag? 
\item Model fit quality: How well do the {\tt JAVELIN} and {\tt CREAM} model light curves match the observed light curve? Are the two model light curves in agreement with one another? 
\item Bootstrapping Results: What is the fraction of consistent samples from the bootstrapping described above? If enough samples are inconsistent with our original lag measurement, this indicates that the lag is less reliable and the object is given a lower quality rating. 

\end{enumerate} 

We include our quality assessments for each lag measurement in Table~\ref{Table:lags}. We recognize that these are subjective --- however, they are based on our significant past experience with RM measurements and thus we provide them to help the reader evaluate the results.

\section{RESULTS AND DISCUSSION}
\label{sec:discussion} 
\subsection{Lag Results} 
\label{sec:lags} 
\begin{figure}
\begin{center}
\includegraphics[scale = 0.45, angle = 0, trim = 0 0 0 0, clip]{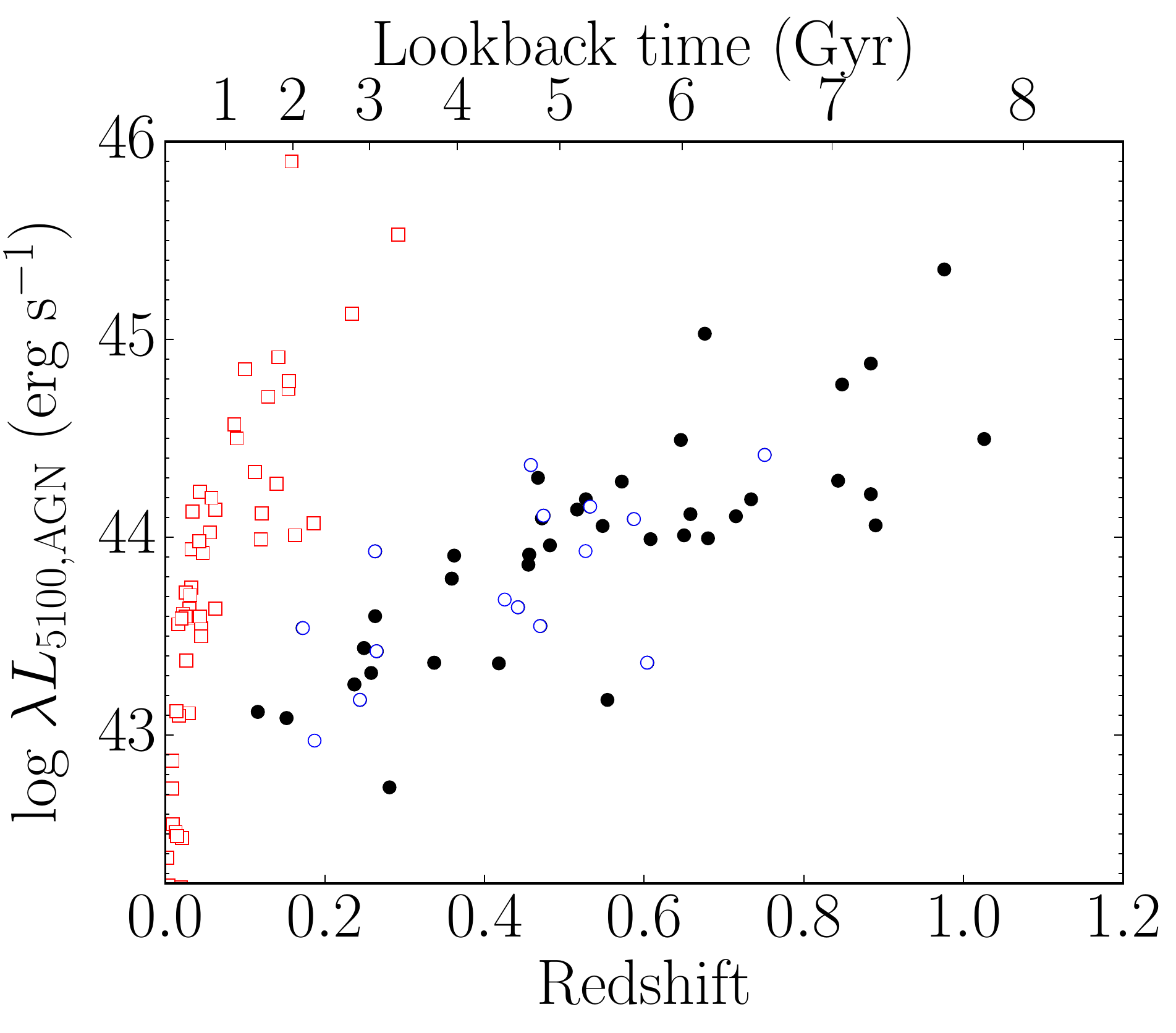}
\caption{The distribution of redshift and log $\lambda L_{5100}$ of the sources with detected \Hbeta \ lags. Red open squares represent the 42 local RM AGN compiled by \cite{Bentz15}, with additions from \cite{Du14, Du15} and \cite{Fausnaugh17}, showing average luminosities when multiple measurements exist for a single source. Blue open circles show the SDSS-RM first-lags sample from \cite{Shen16}, and black closed circles show our new measurements. Note that the \cite{Shen16} measurements include \mgii \ detections and that there is overlap between eight of the \cite{Shen16} \Hbeta \ measurements and our new measurements.}
\label{fig:lum_z_detected} 
\end{center}
\end{figure}

Inspection of the light curves and posterior distributions of sources with lags that were detected by {\tt CREAM} and not {\tt JAVELIN} reveals that {\tt JAVELIN} has a tendency to find more aliases than {\tt CREAM}, particularly in light curves with a longer-term monotonic trend present in the light curve. Despite our alias-removal procedure, the presence of these aliases can cause the measurement to fail our significance criteria despite {\tt JAVELIN} having measured a similar lag as {\tt CREAM}. For our final $\tau$ measurements, we thus adopt $\tau_{\rm JAV}$ if the lag was detected by {\tt JAVELIN} and $\tau_{\rm CREAM}$ for the quasars in which the lag was detected by {\tt CREAM} but not {\tt JAVELIN}. We hereafter refer to the final adopted $\tau$ (which is equivalent to either $\tau_{\rm JAVELIN}$ or $\tau_{\rm CREAM}$) as $\tau_{\rm final}$. 
This procedure yields {\color{black} 32} \Hbeta \ lags from {\tt JAVELIN} alone, and we add {\color{black} 12} more \Hbeta \ lags from {\tt CREAM}, yielding a total of {\color{black} 44} \Hbeta \ lags. Based on the \Hbeta \ false-positive rates estimated for each method (see Sections~\ref{sec:significance} and \ref{sec:methodcomp}), we expect two false positives among the {\tt JAVELIN} lags and two false positives among the {\tt CREAM} lags, yielding an overall number of expected false-positives of 4 out of 44 measurements ($9.1^{+5.6}_{-1.9}$\%).  In addition, we measured {\color{black} 13} \Halpha \ lags from {\tt JAVELIN} and add {\color{black} 5} \Halpha \ lags from {\tt CREAM}, yielding {\color{black} 18} total \Halpha \ lag measurements. Based off of the \Halpha \ false-positive rate, we expect one false positive from {\tt JAVELIN} and less than one from {\tt CREAM}, yielding an expected 1.59 out of 18 \Halpha \ lags ($8.8^{+10.7}_{-2.2}$\%). We provide rest-frame $\tau_{\rm final}$ measurements for all sources with detected lags in either \Hbeta \ or \Halpha \ in Tables~\ref{Table:mbh_hb} and \ref{Table:mbh_ha} and show the luminosity-redshift distribution of these sources in Figure~\ref{fig:lum_z_detected}. We have expanded the redshift range of the RM sample out to $z\sim1$ and increased the number of lag measurements in the sample by {\color{black} about two thirds}.

\cite{Shen16}, hereafter S16, report nine \Hbeta \ lags from the SDSS-RM sample measured from only the spectroscopic light curves. We detect eight of them here and provide the original measurements from S16 (denoted $\tau_{\rm S16}$ and corrected to the observed frame) in Table~\ref{Table:lags} for comparison. Our measurements for the eight detections are all consistent with theirs, but with lower uncertainties due to our addition of the photometric light curves (see Table~\ref{Table:lags}). We find a significantly lower lag for RM 191; this is likely because of the increased cadence of our continuum light curves when the photometric monitoring was incorporated. Because of the increased cadence, we are sensitive to shorter lags and thus are able to measure the shorter lag in this object.  The only source detected by S16 that we do not detect a lag for is RM 769. In our case, all three methods yielded lags that were positive but formally consistent with zero to within the uncertainties. Again, the increased cadence of the light curves is responsible for the difference, allowing us to see that the lag is not well-constrained for this source.  

\begin{figure}
\begin{center}
\includegraphics[scale = 0.3, angle = 0, trim = 0 0 0 0, clip]{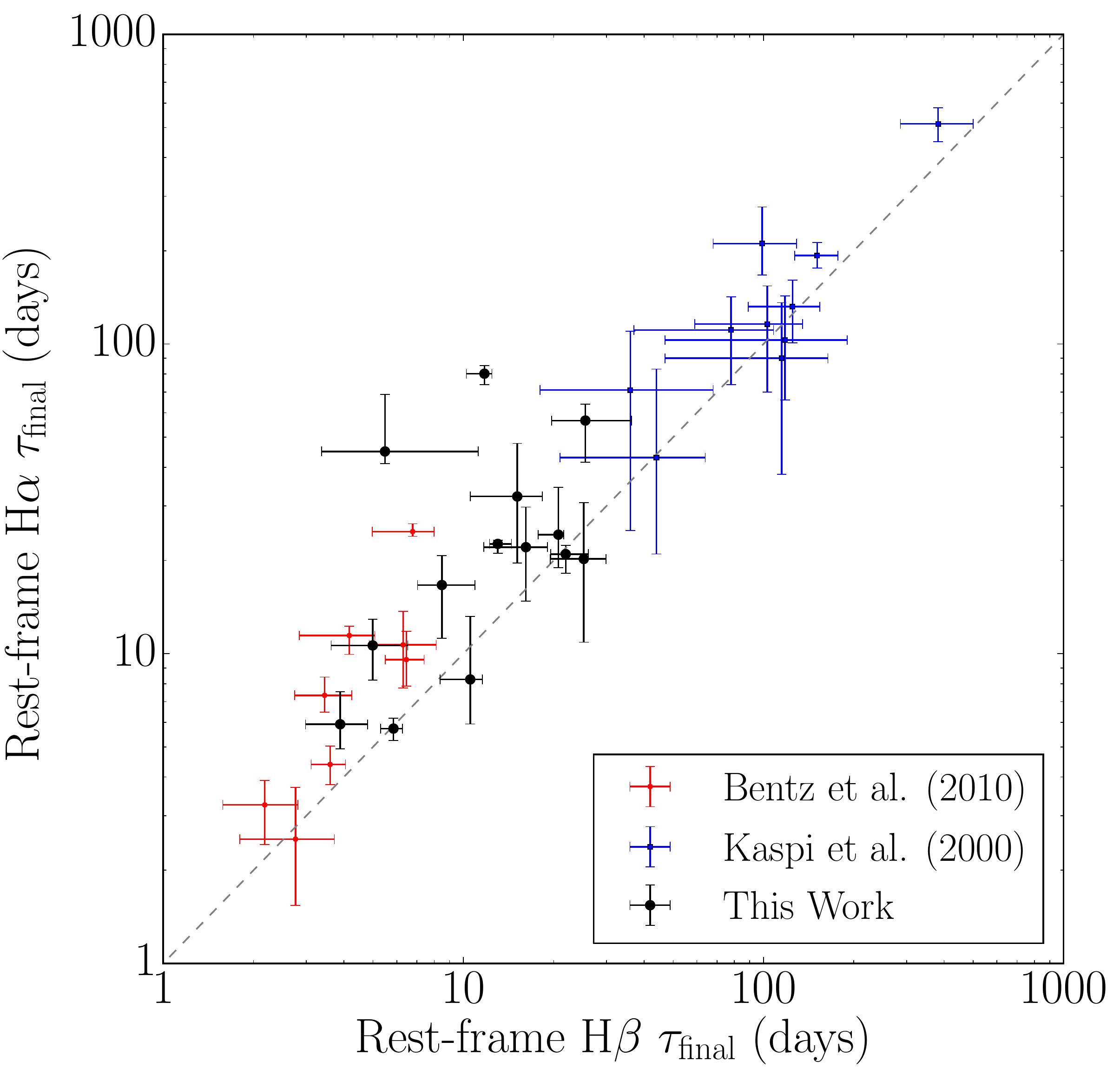}
\caption{\Halpha \ vs. \Hbeta \ lag measurements for those objects where we detected significant lags for both emission lines (black filled circles). Red points represent measurements from \cite{Bentz10a}, and blue squares represent measurements from \cite{Kaspi00}. The gray dashed line shows a ratio of one-to-one to guide the eye. }
\label{fig:ha_vs_hb}
\end{center}
\end{figure}

In {\color{black} 14} quasars, we measure significant lags for both \Hbeta \ and \Halpha \ --- Figure~\ref{fig:ha_vs_hb} compares the \Hbeta \ and \Halpha \ lags for those objects. We see that in all cases, the \Halpha \ lag is consistent with or larger than the \Hbeta \ lag --- this was also reported in previous studies (e.g., \citealt{Kaspi00}; \citealt{Bentz10a}). Larger \Halpha \ lags are expected due to {photoionization predictions, with} radial stratification and optical-depth effects causing the \Halpha \ emission line to appear at larger distances than \Hbeta \ (\citealt{Netzer75}; \citealt{Rees89}; \citealt{Korista04}); see Section 4.3 of \cite{Bentz10a} for a more detailed discussion of this phenomenon. 

\cite{Shen15b} computed the average 5100 \AA \ luminosity of most of our sources during the same monitoring period using spectral decomposition to remove host-galaxy light, allowing us to place these sources on the $R-L$ relation; we provide these luminosities in Table~\ref{Table:sample}. Figure~\ref{fig:rlum} presents the $R-L$ relationship measured by \cite{Bentz13} and shows the location of our new \Hbeta \ lag measurements. {Figure~\ref{fig:rlum} also shows previous RM data from \citet{Du16b} and the compilation of \citet{Bentz15}. For a consistent comparison with our SDSS-RM measurements, we use {\tt JAVELIN} lags when available from the \citet{Bentz15} database. { \color{black} Many of the lags (including the \citealp{Du16b} data) were measured with the ICCF and so typically have larger uncertainties than {\tt JAVELIN} measurements --- however, the lag values themselves are consistent with ICCF measurements and thus there are no issues when comparing measurements made with the various methods. Differences in our lag-measuring procedure (such as adopting the median of the CCCD) also yield measurements that are consistent with those using previously-favored procedures, and thus these lag measurements can also be compared to lags from prior studies without issue. }

{Both our data and the \citet{Du16b} super-Eddington accreting massive black holes (SEAMBH) sample have many AGNs that lie below the $R-L$ relation and its expected scatter. A similar offset from the expected $R-L$ relation was measured for the SDSS-RM quasars using composite cross correlation methods (\citealt{Li17}). At least some of the disagreement may be due to selection effects: the SDSS-RM 2014 cadence and monitoring duration limits our lag detections to less than $\sim$100 days in the observed frame, and it is more difficult to measure the longer lags even below this limit, so we are less likely to measure lags that scatter above the $R-L$ relation. (The  observations had similar cadence and duration.)}

{It is also possible that this offset is due to physical dependencies in the $R-L$ relation. Both the SDSS-RM and SEAMBH quasars { \color{black} lie at the mid-to-high-luminosity end of the $L$ distribution of the \cite{Bentz15} sample of RM quasars}, and it is possible that luminous quasars have different BLR radii than expected from the $R-L$ relation established from low-luminosity AGN. \citet{Du16b} argue that the offset is caused by high accretion rates, since the most rapidly accreting SEAMBH quasars tend to be more frequently offset. We tested this hypothesis by calculating accretion rate using the same parameterization as \citet[][their Equation 3]{Du16b}. In general our SDSS-RM quasars have much (10-1000$\times$) lower accretion rates than the \citet{Du16b} sample (although our quasars have similar $L$ and $R$, they have broader line widths than the narrow-line type 1 AGNs in the SEAMBH sample). The SDSS-RM sample also does not show a clear trend between $R-L$ offset and accretion rate. It is possible that the $R-L$ offset is driven by luminosity rather than accretion rate, or by other quasar properties in which the previous RM samples were biased (e.g., \citealt{Shen15}). Fully exploring the deviations from the $R-L$ relationship will require the multi-year SDSS-RM data and/or careful simulations of the observational biases in order to rule out selection effects. We thus defer more detailed discussion of the $R-L$ relation to future work.}

\begin{figure}
\begin{center}
\includegraphics[scale = 0.31, angle = 0, trim = 0 0 0 0, clip]{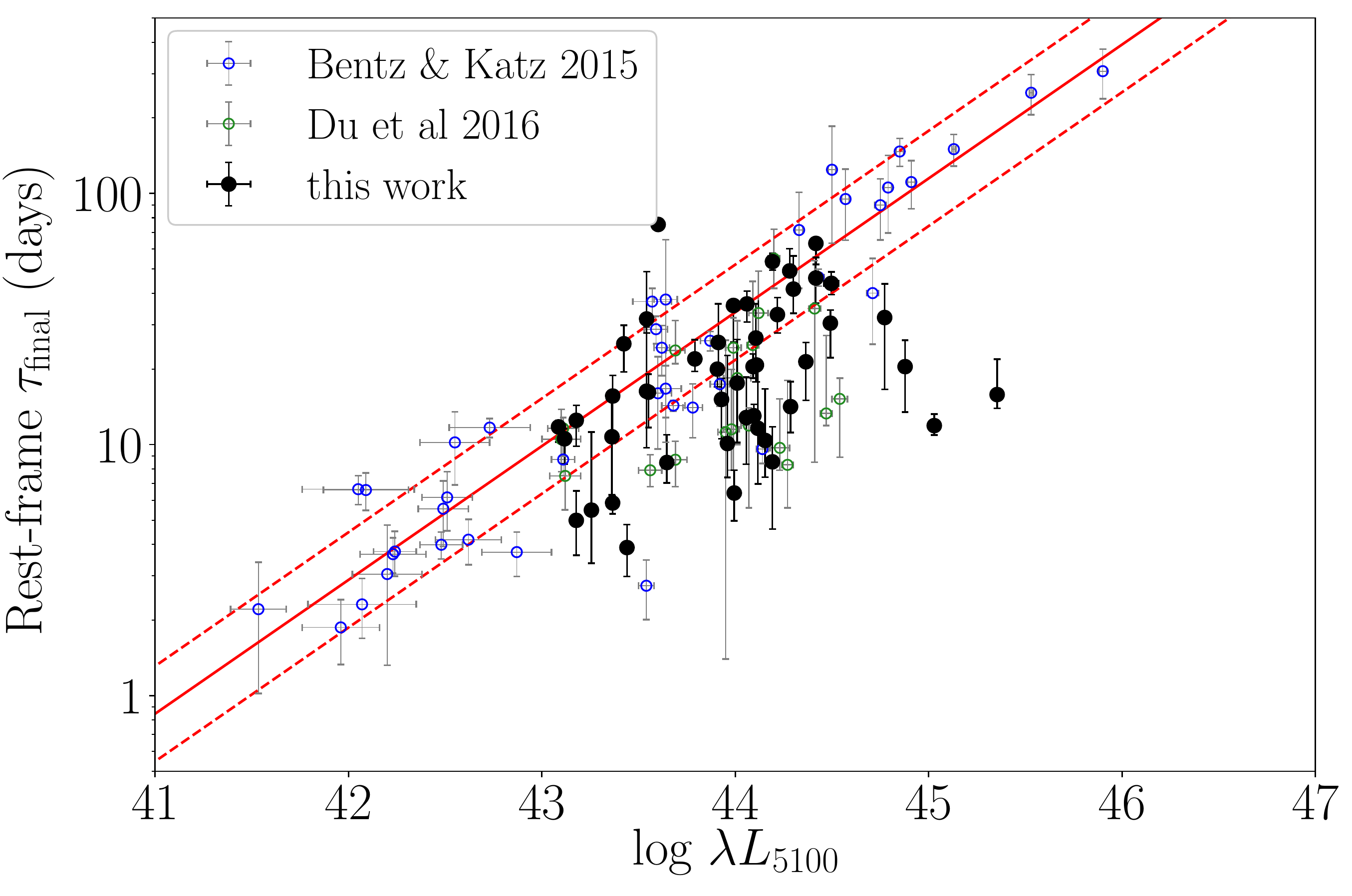}
\caption{{The \Hbeta \ $R-L$ relationship, with previous measurements in blue (\citealp{Bentz15}) and green (\citealp{Du16b}) and our new measurements in black. The red solid and dashed lines show the best-fit relation and its measured scatter from \cite{Bentz13}. Many of the SDSS-RM and \citet{Du16b} lags lie below the main $R-L$ relation: this may be (at least partly) due to selection effects from our limited monitoring cadence and duration, since our survey (and that of \citealp{Du16b}) is not sensitive to long lags at high luminosities. The deviation may also be a physical effect associated with a different BLR size at high luminosities, or other quasar parameters that differ between the initial \citet{Bentz13} dataset and the SDSS-RM data and \citet{Du16b} samples.}}
\label{fig:rlum}
\end{center}
\end{figure}

Our full sample contains 222 quasars; we have thus been able to detect lags in about 20\% of them. Typical yields for traditional RM campaigns with single-object spectrographs (e.g., \citealt{Fausnaugh17}) are on the order of 50\% --- failure in such campaigns, which obtain very high-quality data at high cadences, is usually attributed to a lack of favorable variability behavior of the quasars. These campaigns achieve this 50\% fraction through object selection (the AGN are chosen to have strong emission lines and often are already known to show strong variability), high observing cadence (usually once per day), and high-SNR spectra. Our sample is more representative of quasars with a variety of emission-line properties and luminosities; we thus do not expect as many of our sources to vary in a favorable manner (short-term, high-amplitude variations) during the campaign. In addition, our sample is much fainter on average, which makes flux variations more difficult to detect. The cadence and length of the campaign also affect the yield; we are unable to detect lags longer than $\sim$100 days in the observed frame, which means that lags for the higher-luminosity quasars in our sample (expected to have \Hbeta \ time lags of up to $\sim$300 days in the observed frame), are undetectable with this dataset. We expect that future programs similar to SDSS-RM will similarly yield a $\sim$20\% detection fraction over the first year (although the fraction may be higher for a brighter subset of quasars), with improvements if the overall cadence and monitoring length are increased. 

\subsection{Black-Hole Mass Measurements} 
\label{sec:mbh} 
We use our $\tau_{\rm final}$ measurements in combination with line-width measurements from PrepSpec to compute \mbh for our sources following Equation~1. We report these line-width measurements, along with the adopted lags, calculated virial products, and \mbh measurements for \Hbeta \ in Table~\ref{Table:mbh_hb}, and  \Halpha \ in Table~\ref{Table:mbh_ha}. To calculate the virial products, we use $\sigma_{\rm line, rms}$ measured from the RMS residual spectrum, which has been shown to be a less biased estimator for \mbh than the FWHM for \Hbeta-based measurements (\citealt{Peterson11}). We note that the PrepSpec RMS spectrum is different from ``traditional" RMS spectra used in many previous studies (e.g., \citealt{Kaspi00}; \citealt{Peterson04}). Most prior studies include the entire spectrum, including the continuum and any blended components, in the RMS spectrum computation. PrepSpec decomposes the spectra into multiple components, and the RMS line profiles are measured from the broad-line model only. The resulting RMS widths are different from those measured from the entire spectrum --- \cite{Barth15} examined possible sources of systematics in the RMS line-width measurements and found that the inclusion of the continuum in the RMS calculation can cause the line widths to be underestimated (see \citealt{Barth15}, Appendix C, for details). 

We propagate the uncertainties in $\tau_{\rm JAV}$ and $\sigma_{\rm line}$ to compute the statistical uncertainties on the virial product; however, there are additional systematic uncertainties affecting \mbh measurements that have not yet been taken into account. \cite{Fausnaugh17} calculate a 0.16 dex standard deviation in the mass of the BH in NGC\,5548, which has been measured by many independent monitoring campaigns over the past 30 years (\citealt{Bentz15}). We follow \cite{Fausnaugh17} and add 0.16 dex uncertainties in quadrature with the statistical uncertainties in the virial product to produce our final adopted uncertainties. We adopt a scale factor $f= 4.47$ (\citealt{Woo15}) to convert the virial products to $M_{\rm BH}$.  We note that the 0.16 dex systematic uncertainty is negligible compared to the systematic uncertainty in the virial scale factor $f$ (generally recognized to be on the order of 0.4 dex). 

We also compare our \mbh measurements ($M_{\rm BH, RM}$) from the \Hbeta \ emission line with those measured by \cite{Shen15b} using single-epoch spectra and the prescription of \cite{Vestergaard06}, { \color{black} hereafter VP06,} for objects with 5100\,\AA \ luminosity measurements ($M_{\rm BH, SE}$). Before comparing measurements, however, we increased the reported statistical uncertainties of the single-epoch measurements by 0.43 dex to account for the intrinsic scatter measured by VP06 in the single-epoch \mbh \ calibration. { \color{black} VP06 used a higher scaling factor than our adopted value, which results in slightly higher (by 0.1~dex) single-epoch masses (VP06 adopt $f_{\sigma}=5.5$ and $f_{\rm FWHM}=1.4$ from \citealt{Onken04}, while we adopt $f_{\sigma}=4.47$ and $f_{\rm FWHM}=1.12$ from \citealt{Woo15}). 

Figure~\ref{fig:se_vs_rm} shows a comparison between the two measurements. In most cases, our \mbh\ measurements are consistent with the single-epoch measurements given the uncertainties. The agreement is even better if a correction is applied for the different scaling factor. The scatter around a 1-to-1 relation among our sample is similar to the scatter seen among the the \cite{Bentz15} and \cite{Du16b} samples.} However, both our sample and that of \citet{Du16b} have slightly over massive single-epoch \mbh at low RM masses. This is consistent with the deviation seen from the $R-L$ relation (Figure \ref{fig:rlum}), with smaller RM lags than expected from luminosity and the \citet{Bentz13} relation. As before, it is possible that the differences are associated with differences in quasar properties: our sources are more luminous than those of \citet{Bentz13}, though not as rapidly accreting as the \citet{Du16b} quasars. However, it is also possible that the apparent deviation is caused by selection effects associated with our limited cadence and duration, and so we withhold definitive conclusions until detailed simulations of the observational biases are examined in future work.

We also compare our \mbh measurements from \Hbeta\ with those from \Halpha\ in Figure~\ref{fig:mbh_ha_vs_hb}, and find that the measurements are consistent to within the uncertainties for nearly all sources.
\begin{figure}
\begin{center}
\includegraphics[scale = 0.37, angle = 0, trim = 0 0 0 0, clip]{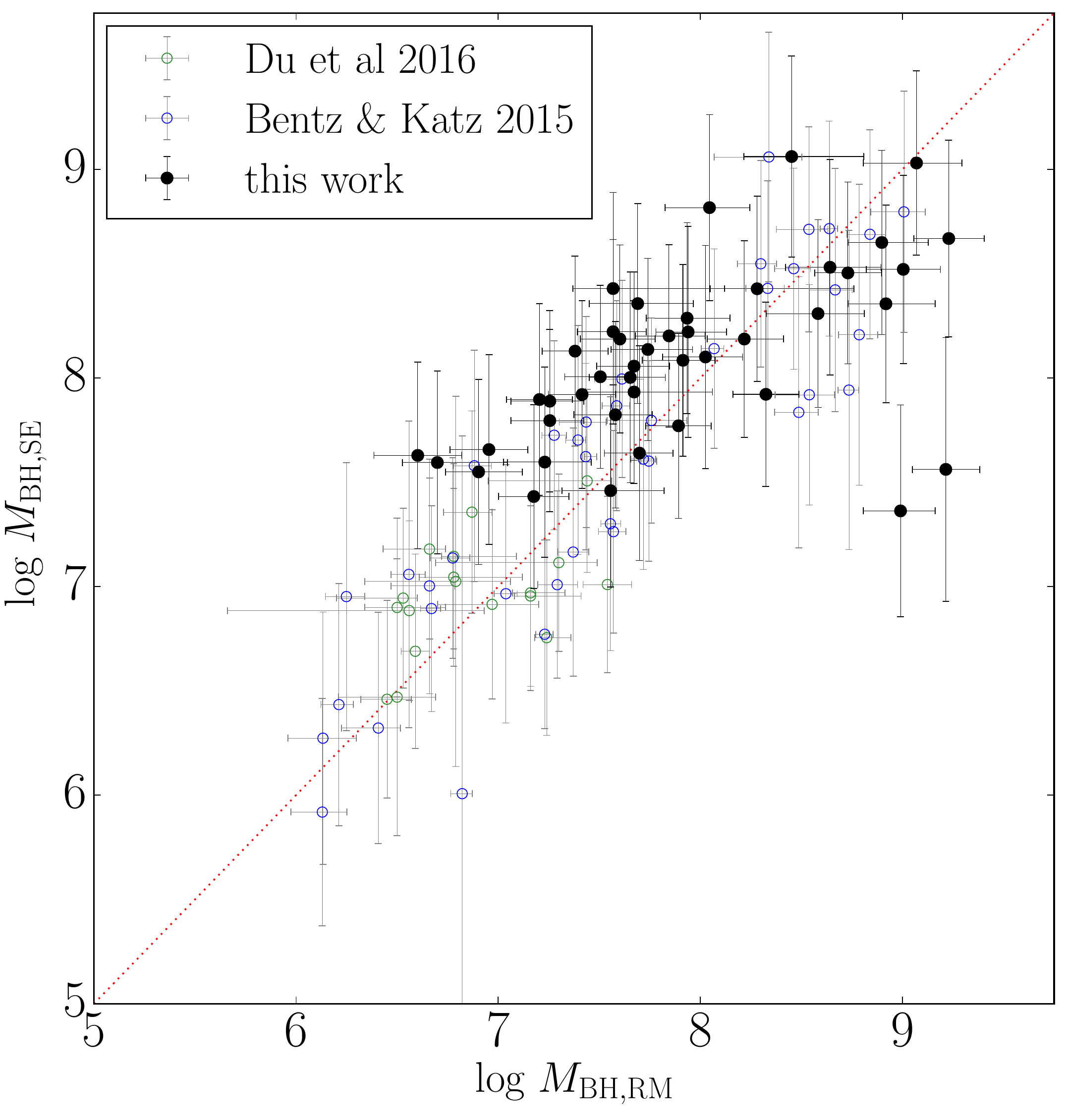}
\caption{{Comparison of $M_{\rm BH, SE}$ measurements from \cite{Shen15b} and our new measurements ($M_{\rm BH, RM}$). As in Figure \ref{fig:rlum}, we show previous measurements in blue (\citealp{Bentz15}) and green (\citealp{Du16b}). The dotted red line indicates a ratio of one-to-one. Most of our quasars have consistent masses between the two methods, with some deviation for both SDSS-RM and the \citet{Du16b} sample at low RM masses.}}
\label{fig:se_vs_rm}
\end{center}
\end{figure} 

Figure~\ref{fig:msigma} places our \mbh measurements on the \msigma relationship. These $\sigma_*$ measurements were taken from \cite{Shen15b}, but they are also consistent with those measured independently by \cite{Matsuoka15} based on a different spectral decomposition approach. { \color{black} Most of our measurements are consistent with the local quiescent \msigma relation, though large uncertainties and the presence of outliers at low-$\sigma_*$ introduce a large amount of scatter and dilute any correlation within our sample}. 
{ \color{black} The four outliers at low-$\sigma_*$ are RMIDs 320, 338, 392, and 779. All four of these lag measurements appear solid --- we see visible short-term variability in the light curves and the lags are well-determined, with clean posteriors. In addition, three out of the four lags are consistent with expectations from the $R-L$ relation, further suggesting that they are robust (the fourth source, RM 392, is expected to have a much longer lag than we measure, however). Upon inspection we find that there are likely isues with the $\sigma_*$ measurements, all of which are below 100~\kms\ and approach the limits of the data used to measure them. 
We examined the spectral decomposition fits used to measure $\sigma_*$ in these four sources and found that using the Ca H/K lines only, the measurements for these sources are significantly higher, indicating that the original measurements are likely underestimated; this is what causes them to appear as outliers on the \msigma relation.}

\begin{figure}
\begin{center}
\includegraphics[scale = 0.3, angle = 0, trim = 0 0 0 0, clip]{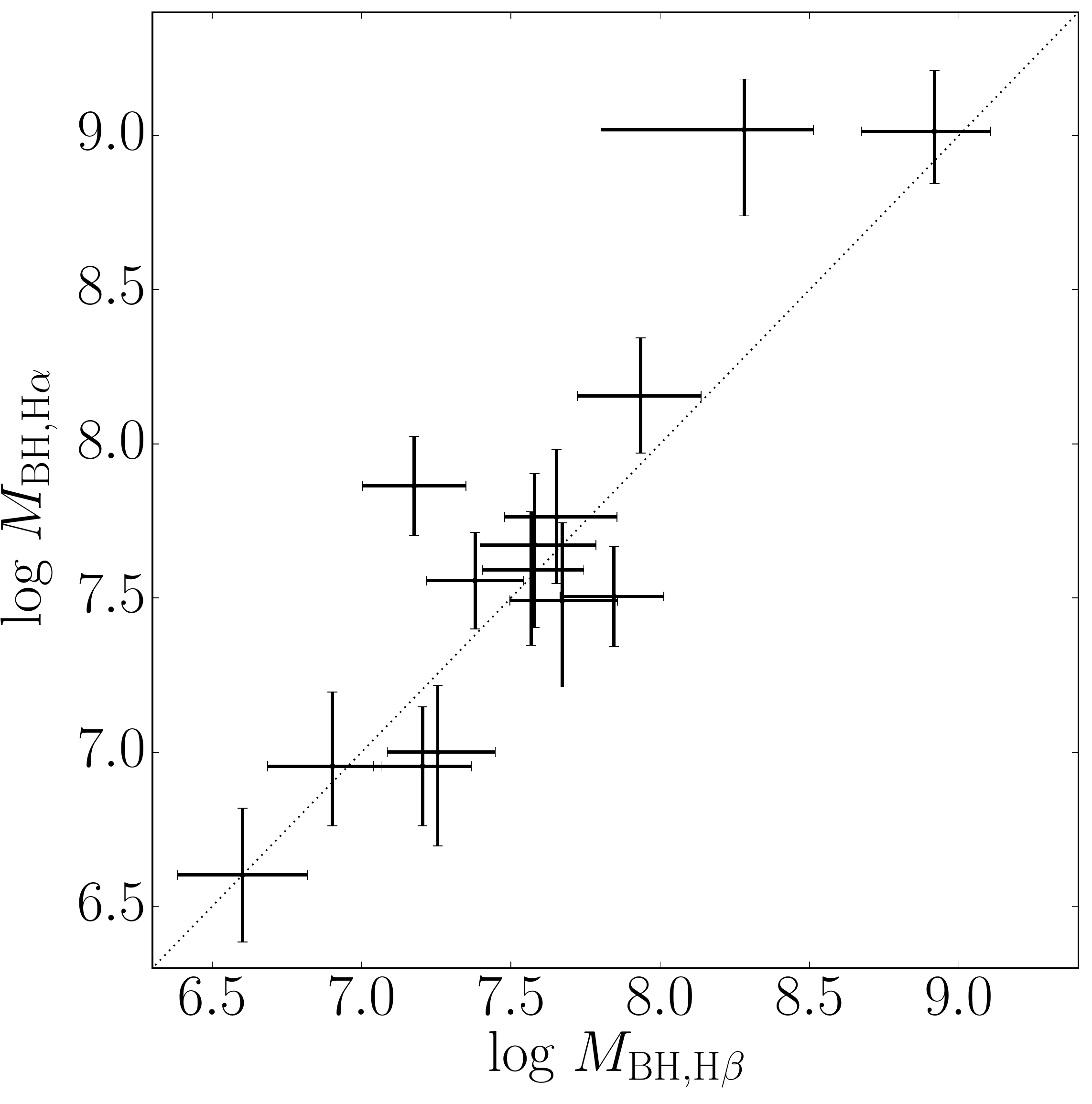}
\caption{Comparison of $M_{\rm BH}$ measured from \Hbeta \ and \Halpha \ for those objects where we detected lags in both emission lines. The black dotted line shows a ratio of one-to-one. }
\label{fig:mbh_ha_vs_hb}
\end{center}
\end{figure}

\begin{figure}
\begin{center}
\includegraphics[scale = 0.43, angle = 0, trim = 0 0 0 0, clip]{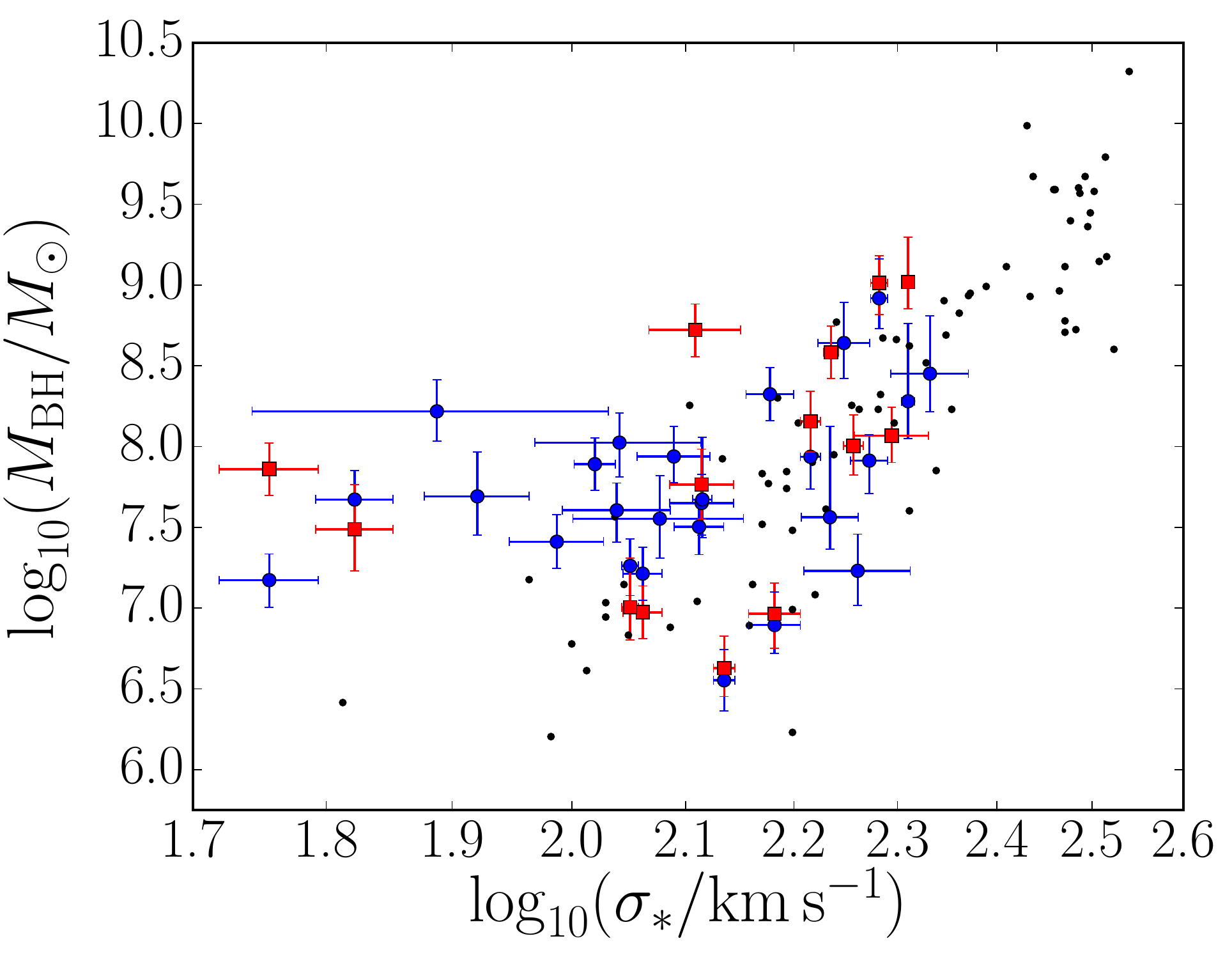}
\caption{The \msigma \ relation with the sample of dynamical black hole masses from \cite{McConnell13} shown as black dots. Our new \mbh measurements made using the \Hbeta \ and \Halpha \ emission line time lags and line widths are represented by blue circles and red squares, respectively.}
\label{fig:msigma}
\end{center}
\end{figure}

\subsection{Additional Sources of Systematics} 
\label{sec:systematics} 
When obtaining RM $R_{\rm BLR}$ measurements, RM studies often make the assumption that the time delay between the 5100\,\AA \ continuum-emitting region and a relevant broad emission line (such as \Hbeta) is a good characterization of the distance between the BH and the \Hbeta-emitting BLR. However, the $R_{\rm BLR}$ measured with RM is actually the distance between the optical continuum-emitting region and the BLR and not between the BLR and the BH itself; past RM efforts have generally assumed that the distance between the BH and the continuum-emitting region is negligible. Recent work has indicated that the optical continuum-emitting region can have a significant lag relative to the UV (\citealt{Collier98}; \citealt{Sergeev05}; \citealt{Mchardy14}; \citealt{Shappee14}; \citealt{Edelson15}; \citealt{Fausnaugh16a}). Specifically, \cite{Fausnaugh16a} found that the $V$-band emitting region of NGC\,5548 is 2 light-days farther out than the 1367\,\AA \ UV-emitting region in NGC\,5548, a distance that is non-negligible compared to the measured time lags for some of the broad emission lines (the He\,{\sc ii} $\lambda1640$ emission line was measured to have a time delay of 2.4 days, for example). It is thus likely that our measurements of $R_{\rm BLR}$ using optical/BLR lags are underestimated. 

Assuming a universal AGN accretion disk, where the distance between the UV and optical continuum-emitting regions is constant for all sources, a nonzero UV-optical time delay will not have an effect on RM \mbh measurements because the scale factor $f$ automatically corrects, at least in a statistical sense, for this distance effect by requiring that AGN fall on the quiescent \msigma relation. However, if the UV/optical distance depends on $L$ and/or $M_{\rm BH}$, this complication could pose a problem.  
\cite{Pei17} examine possible dependencies on quasar parameters such as $L$ and \mbh and report that the scaling with luminosity is expected to be slow. Microlensing studies also show that the size of the BLR more or less scales as expected with \mbh (\citealt{Morgan10}; \citealt{Mosquera13}).

In addition, the scatter in the $R-L$ relationship (\citealt{Bentz13}) is small --- thus these effects are likely small for most AGN, as a large UV/optical distance scaling would cause larger scatter in this relation. Thus far, there are only a few solid measurements of the UV-optical continuum time delay, so we are unable to directly measure any dependencies of UV-optical size with quasar properties. Additional measurements of inter-band continuum lags will be necessary to determine what (if any) correction is needed to account for the use of optical continuua in measuring broad emission-line lags.

\section{Summary}
\label{sec:summary} 
We have combined the spectroscopic and photometric observations from the first year of monitoring of the SDSS-RM program to search for significant time delays in 222 quasars. Our major findings are the following: 
\begin{enumerate} 
\item We have measured characteristic time delays between the continuum and the \Hbeta \ and \Halpha \ broad emission lines in {\color{black} 44} and {\color{black} 18} sources, respectively. These measurements increase the size of the sample of AGN that have reverberation mapping \mbh measurements by about {\color{black} two thirds}. In addition, most of these measurements are made for higher-redshift objects, significantly expanding the redshift coverage of the RM sample. See Section~\ref{sec:discussion}. 
\item We compared three different methods of obtaining lag measurements: the ICCF, {\tt JAVELIN}, and {\tt CREAM}. All three methods are generally consistent with one another, though {{\tt JAVELIN} (32 \Hbeta\ and 13 \Halpha\ lags) and {\tt CREAM} (42 \Hbeta\ and 17 \Halpha\ lags) typical yield smaller uncertainties and thus more high-significance detections than the ICCF (16 \Hbeta\ and 8 \Halpha\ lags)}. See Section~\ref{sec:methodcomp}. 
\item We find that \Halpha \ lags are generally consistent with or larger than the \Hbeta \ lags measured in the same sources, which is consistent with previous findings. See Section~\ref{sec:lags}. 
\item We find that many of our sources fall below the $R-L$ relation measured by \cite{Bentz13}. This could be due to selection effects or a dependency of the $R-L$ relation on accretion parameters such as the Eddington ratio. See Section~\ref{sec:lags}. 
\item We measure \mbh for those objects with successful lag detections. Most of our measurements are consistent with the \msigma relation measured in local quiescent galaxies, though we have some outliers at the low-$\sigma_*$ end of the relation that are likely due to selection effects and/or issues with $\sigma_*$ measurements. See Section~\ref{sec:mbh}. 
\end{enumerate} 

With only the first year of data, we are sensitive only to lag measurements shorter than approximately 100 days in the observed frame. The next step is to incorporate the additional years of data from the SDSS-RM program to extend the lag sensitivity and the dynamic range in quasar luminosity. This will allow us to measure longer time delays and also will help remove aliases in our posterior lag distributions for shorter lags, which will likely reduce our false-positive rate.  With the additional years of data that are already in hand or have been planned, we will also be able to investigate emission lines such as  \civ \ and \mgii \ in higher-redshift targets, allowing us to probe quasars even further out in the Universe. 

\acknowledgments CJG, WNB, JRT, and DPS acknowledge support from NSF grant AST-1517113. YS acknowledges support from an Alfred P. Sloan Research Fellowship and NSF grant AST-1715579. CSK is supported by NSF grant AST-1515427. KH acknowledges support from STFC grant ST/M001296/1. IDM acknowledges support from NSF grant AST 15-15115. KDD acknowledges support from the NSF awarded under NSF Grant AST-1302093. LCH was supported by the National Key R\&D Program of China (2016YFA0400702) and the National Science Foundation of China (11473002, 11721303).

This work is based on observations obtained with MegaPrime/MegaCam, a joint project of CFHT and CEA/DAPNIA, at the Canada-France-Hawaii Telescope (CFHT) which is operated by the National Research Council (NRC) of Canada, the Institut National des Sciences de l'Univers of the Centre National de la Recherche Scientifique of France, and the University of Hawaii.  The authors wish to recognize and acknowledge the very significant cultural role and reverence that the summit of Maunakea has always had within the indigenous Hawaiian community.  The astronomical community is most fortunate to have the opportunity to conduct observations from this mountain.

Funding for SDSS-III has been provided by the Alfred P. Sloan Foundation, the Participating Institutions, the National Science Foundation, and the U.S. Department of Energy Office of Science. The SDSS-III web site is http://www.sdss3.org/.

SDSS-III is managed by the Astrophysical Research Consortium for the Participating Institutions of the SDSS-III Collaboration including the University of Arizona, the Brazilian Participation Group, Brookhaven National Laboratory, Carnegie Mellon University, University of Florida, the French Participation Group, the German Participation Group, Harvard University, the Instituto de Astrofisica de Canarias, the Michigan State/Notre Dame/JINA Participation Group, Johns Hopkins University, Lawrence Berkeley National Laboratory, Max Planck Institute for Astrophysics, Max Planck Institute for Extraterrestrial Physics, New Mexico State University, New York University, Ohio State University, Pennsylvania State University, University of Portsmouth, Princeton University, the Spanish Participation Group, University of Tokyo, University of Utah, Vanderbilt University, University of Virginia, University of Washington, and Yale University.

We thank the Bok and CFHT Canadian, Chinese, and French TACs for their support. This research uses data obtained through the Telescope Access Program (TAP), which is funded by the National Astronomical Observatories, Chinese Academy of Sciences, and the Special Fund for Astronomy from the Ministry of Finance in China.

\clearpage
\LongTables  
\begin{deluxetable*}{lcccccrccc}
\tablewidth{0pt}
\tablecaption{Quasar Sample Information}
\tablehead{ 
\colhead{} & 
\colhead{} & 
\colhead{RA\tablenotemark{a}} & 
\colhead{Dec\tablenotemark{a}} & 
\colhead{} & 
\colhead{} & 
\colhead{} & 
\colhead{AGN} & 
\colhead{host} & 
\colhead{} \\
\colhead{} & 
\colhead{SDSS} & 
\colhead{(deg)} & 
\colhead{(deg)} & 
\colhead{} & 
\colhead{} & 
\colhead{MED\tablenotemark{b}} & 
\colhead{log $\lambda L_{\lambda 5100}$\tablenotemark{c}} & 
\colhead{log $\lambda L_{\lambda 5100}$\tablenotemark{c}} & 
\colhead{log $M_{\rm BH, SE}$\tablenotemark{c}} \\
\colhead{RMID} & 
\colhead{Identifier} & 
\colhead{(J2000)} & 
\colhead{(J2000)} & 
\colhead{$z$\tablenotemark{a}} & 
\colhead{$i$ mag\tablenotemark{a}} & 
\colhead{SNR} & 
\colhead{(erg s$^{-1}$)} & 
\colhead{(erg s$^{-1}$)} & 
\colhead{(\Msun)}
}
\startdata
005 & J141541.41+530424.3 & 213.9225 &  53.0734 &  1.020 & 20.716 &  5.0 & 44.5 & \nodata &  8.1 \\
009 & J141359.51+531049.3 & 213.4980 &  53.1804 &  0.898 & 20.473 &  4.3 & 44.1 & 43.8 &  7.9 \\
016 & J141606.95+530929.8 & 214.0290 &  53.1583 &  0.848 & 19.716 & 11.5 & 44.8 & 43.9 &  9.0 \\
017 & J141324.28+530527.0 & 213.3511 &  53.0908 &  0.456 & 19.213 & 14.1 & 43.9 & 44.2 &  8.4 \\
018 & J141323.27+531034.3 & 213.3469 &  53.1762 &  0.849 & 20.205 &  5.2 & 44.3 & 44.2 &  8.9 \\
\smallskip
020 & J141411.66+525149.0 & 213.5486 &  52.8636 &  1.124 & 21.529 &  2.0 & \nodata & \nodata & \nodata \\
021 & J141314.97+530139.4 & 213.3124 &  53.0276 &  1.026 & 21.167 &  3.0 & 44.5 & 44.0 &  7.6 \\
027 & J141600.80+525255.5 & 214.0033 &  52.8821 &  1.023 & 20.968 &  2.1 & 44.4 & 44.2 &  8.5 \\
\enddata
\label{Table:sample}
\tablenotetext{a}{These measurements were made as a part of the SDSS Data Release 10 (\citealt{Ahn14}). The $i$-magnitudes listed are PSF magnitudes, and have not been corrected for Galactic extinction. }
\tablenotetext{b}{MED-SNR, where SNR is the median signal-to-noise ratio per SDSS pixel across each individual spectrum, and MED-SNR is the median across all epochs (each SDSS pixel spans 69 km s$^{-1}$).}
\tablenotetext{c}{These measurements are taken from \cite{Shen15b}. $M_{\rm BH, SE}$ estimates were made using the \cite{Vestergaard06} prescription for $L_{5100}$.}
\end{deluxetable*}

\begin{deluxetable*}{lccrrc} 
\tablewidth{0pt} 
\tablecaption{RM 005 Light Curves } 
\tablehead{ 
\colhead{MJD} & 
\colhead{} & 
\colhead{} & 
\colhead{} & 
\colhead{} & 
\colhead{} \\ 
\colhead{(-50000)} & 
\colhead{Band\tablenotemark{a}} & 
\colhead{Telescope\tablenotemark{b}} & 
\colhead{Flux\tablenotemark{c}} & 
\colhead{Error\tablenotemark{c}} & 
\colhead{FLAG\tablenotemark{d}}  
}  
\startdata 
      6660.2090 & $g$ & S &      16.99  &       0.33 &  0   \\ 
      6664.5130 & $g$ & S &      16.99  &       0.33 &  0   \\ 
      6669.5003 & $g$ & S &      16.90  &       0.34 &  0   \\ 
      6686.4734 & $g$ & S &      17.47  &       0.35 &  0   \\ 
      6711.5226 & $g$ & C &      17.48  &       0.19 &  0   \\ 
      6712.4684 & $g$ & C &      17.75  &       0.18 &  0   \\ 
      6712.4694 & $g$ & C &      17.28  &       0.18 &  0   \\ 
      6712.4703 & $g$ & C &      17.48  &       0.19 &  0   \\
\enddata 
\tablecomments{Light curves for all 222 quasars can be found online. A portion is shown here for guidance in formatting.  }
\tablenotetext{a}{\Hbeta \ = \Hbeta \ emission line, \Halpha \ = \Halpha \ emission line, $g$ = $g$-band and $i$ = $i$-band. } 
\tablenotetext{b}{C = CFHT, B = Bok, S = SDSS. } 
\tablenotetext{c}{Continuum Flux densities and uncertainties are in units of 10$^{-17}$ erg s$^{-1}$ cm$^{-2}$ \AA$^{-1}$. Integrated emission-line fluxes are in units of 10$^{-17}$ erg s$^{-1}$ cm$^{-2}$. {The fluxes are not host-subtracted.}} 
\tablenotetext{d}{Emission-line epochs with FLAG = 1 were identified as outliers and excluded from the light curves in our analysis.} 
\label{Table:lc_rm005} 
\end{deluxetable*}  

\clearpage 

\begin{deluxetable*}{lc|rrrc|r|rrrc} 
\tablewidth{0pt} 
\tablecaption{SDSS-RM Observed-Frame Lag Detections} 
\tablehead{ 
\colhead{} & \colhead{} & \multicolumn{5}{c}{\Hbeta \ Results} & \multicolumn{4}{c}{\Halpha \ Results} \\ 
\hline 
\colhead{} & 
\colhead{} & 
\colhead{$\tau_{\rm ICCF}$} & 
\colhead{$\tau_{\rm JAV}$} & 
\colhead{$\tau_{\rm CREAM}$} & 
\colhead{Quality\tablenotemark{a}} & 
\colhead{$\tau_{\rm S16}$\tablenotemark{b}} & 
\colhead{$\tau_{\rm ICCF}$} & 
\colhead{$\tau_{\rm JAV}$} & 
\colhead{$\tau_{\rm CREAM}$} & 
\colhead{Quality\tablenotemark{a}} \\ 
\colhead{RMID} & 
\colhead{$z$} & 
\colhead{(days)} & 
\colhead{(days)} & 
\colhead{(days)} & 
\colhead{Rating} & 
\colhead{(days)} & 
\colhead{(days)} & 
\colhead{(days)} & 
\colhead{(days)} & 
\colhead{Rating}  
}  
\startdata 
016 & 0.848 & {\color{black}$ 55.0^{+ 9.3}_{- 9.2}$} & {\color{black}$ 64.5^{+21.5}_{-34.6}$} & {\color{black} $ 59.2^{+21.4}_{-28.6}$} & 3 & \nodata & \nodata & \nodata & \nodata & \nodata \\ 
017 & 0.456 & {\color{black} $ 32.7^{+15.4}_{-15.9}$} & {\color{black} $ 37.1^{+15.8}_{- 8.5}$} & {\color{black} $ 36.7^{+ 4.3}_{- 8.4}$} & 4 & \nodata & {\color{black} $ 65.3^{+20.0}_{-15.4}$} & {\color{black} $ 82.4^{+10.6}_{-21.9}$} & {\color{black} $ 65.7^{+ 3.6}_{-13.7}$} & 5 \\ 
021 & 1.026 & {\color{black}$-86.0^{+10.0}_{- 5.7}$} & {\color{black} $ 88.9^{+ 9.9}_{- 8.8}$} & {\color{black}$-85.8^{+ 5.0}_{- 3.4}$} & 2 & \nodata & \nodata & \nodata & \nodata & \nodata \\ 
033 & 0.715 & {\color{black}$ 19.0^{+20.4}_{-15.9}$} & {\color{black}$ 47.7^{+11.0}_{- 7.7}$} & {\color{black} $ 45.5^{+16.9}_{-15.1}$} & 1 & \nodata & \nodata & \nodata & \nodata & \nodata \\ 
\smallskip 
088 & 0.516 & \nodata & \nodata & \nodata & \nodata & \nodata & {\color{black} $ 84.0^{+ 5.7}_{- 8.3}$} & {\color{black} $ 83.1^{+ 4.4}_{- 7.7}$} & {\color{black}$-82.0^{+ 3.6}_{- 3.2}$} & 3 \\ 
101 & 0.458 & {\color{black} $ 31.1^{+ 5.9}_{-11.4}$} & {\color{black} $ 31.1^{+ 6.1}_{- 9.3}$} & {\color{black} $ 30.1^{+ 6.8}_{- 8.4}$} & 5 & \nodata & \nodata & \nodata & \nodata & \nodata \\ 
160 & 0.359 & {\color{black}$ 14.6^{+ 8.9}_{- 9.6}$} & {\color{black}$ 31.3^{+ 8.1}_{- 4.1}$} & {\color{black} $ 29.8^{+ 5.7}_{- 3.2}$} & 3 & \nodata & {\color{black}$ 16.5^{+10.3}_{- 9.2}$} & {\color{black}$ 27.7^{+ 5.3}_{- 4.7}$} & {\color{black} $ 28.5^{+ 1.9}_{- 3.8}$} & 4 \\ 
177 & 0.482 & {\color{black}$  8.3^{+ 9.1}_{- 6.6}$} & {\color{black} $ 15.0^{+18.6}_{- 4.0}$} & {\color{black}$ 33.4^{+ 1.7}_{-24.3}$} & 4 & \nodata & \nodata & \nodata & \nodata & \nodata \\
\enddata 
\tablenotetext{a}{Lag quality rating (see Section \ref{sec:quality}). } 
\tablenotetext{b}{The lag measured by \cite{Shen16}, for comparison purposes. } 
\label{Table:lags} 
\end{deluxetable*}  

\clearpage
\begin{deluxetable*}{lccrrrrrrr} 
\tablewidth{0pt} 
\tablecaption{Line Width, Virial Product, and $M_{\rm BH}$ Measurements for \Hbeta} 
\tablehead{ 
\colhead{ } & 
\colhead{ } & 
\colhead{$\tau_{\rm final}$\tablenotemark{a}} & 
\colhead{$\sigma_{\rm line, mean}$\tablenotemark{b}} & 
\colhead{$\sigma_{\rm line, rms}$\tablenotemark{b}} & 
\colhead{FWHM$_{\rm mean}$\tablenotemark{b}} & 
\colhead{FWHM$_{\rm rms}$\tablenotemark{b}} & 
\colhead{VP} & 
\colhead{$M_{\rm BH}$\tablenotemark{c}} & 
\colhead{$\sigma_{*}$\tablenotemark{d}} \\ 
\colhead{RMID} & 
\colhead{$z$} & 
\colhead{(days)} & 
\colhead{(km s$^{-1}$)} & 
\colhead{(km s$^{-1}$)} & 
\colhead{(km s$^{-1}$)} & 
\colhead{(km s$^{-1}$)} & 
\colhead{(10$^7$ \Msun)} & 
\colhead{(10$^7$ \Msun)} & 
\colhead{(km s$^{-1}$)}  
}  
\startdata 
RM 016  &  0.848  &  $ 32.0^{+ 11.6}_{- 15.5}$ & $ 4585 \pm    50$ & $ 6477 \pm    54$ & $ 7976 \pm    42$ & $ 7859 \pm   112$ & $ 26.2^{+ 13.5}_{- 15.9}$ &  $117.3^{+ 60.5}_{- 71.2}$ & \nodata \\ 
RM 017  &  0.456  &  $ 25.5^{+ 10.9}_{-  5.8}$ & $ 6937 \pm    14$ & $ 6101 \pm    48$ & $16318 \pm    30$ & $ 7758 \pm    77$ & $ 18.5^{+ 10.4}_{-  8.0}$ &  $ 82.8^{+ 46.6}_{- 35.9}$ & $191.4 \pm   3.7$ \\ 
RM 021  &  1.026  &  $ 43.9^{+  4.9}_{-  4.3}$ & $ 4856 \pm   536$ & $ 6543 \pm    34$ & $ 7442 \pm  5983$ & $11002 \pm  1743$ & $ 36.6^{+ 14.1}_{- 14.0}$ &  $163.8^{+ 63.1}_{- 62.5}$ & \nodata \\ 
RM 033  &  0.715  &  $ 26.5^{+  9.9}_{-  8.8}$ & $  776 \pm    13$ & $  857 \pm    32$ & $ 1070 \pm    30$ & $ 1626 \pm   243$ & $  0.4^{+  0.2}_{-  0.2}$ &  $  1.7^{+  0.9}_{-  0.8}$ & $182.4 \pm  21.7$ \\ 
\smallskip 
RM 101  &  0.458  &  $ 21.4^{+  4.2}_{-  6.4}$ & $ 1186 \pm     2$ & $  976 \pm    32$ & $ 2220 \pm     9$ & $ 2135 \pm    94$ & $  0.4^{+  0.2}_{-  0.2}$ &  $  1.8^{+  0.7}_{-  0.8}$ & \nodata \\ 
RM 160  &  0.359  &  $ 21.9^{+  4.2}_{-  2.4}$ & $ 1773 \pm     6$ & $ 1909 \pm    12$ & $ 4399 \pm    31$ & $ 4183 \pm    51$ & $  1.6^{+  0.6}_{-  0.6}$ &  $  7.0^{+  2.9}_{-  2.7}$ & \nodata \\ 
RM 177  &  0.482  &  $ 10.1^{+ 12.5}_{-  2.7}$ & $ 2541 \pm     9$ & $ 2036 \pm    39$ & $ 5277 \pm    39$ & $ 4930 \pm   163$ & $  0.8^{+  1.1}_{-  0.4}$ &  $  3.7^{+  4.7}_{-  1.7}$ & $171.5 \pm  10.7$ \\ 
RM 191  &  0.442  &  $  8.5^{+  2.5}_{-  1.4}$ & $  845 \pm    12$ & $ 1030 \pm    18$ & $ 1316 \pm    94$ & $ 1967 \pm    76$ & $  0.2^{+  0.1}_{-  0.1}$ &  $  0.8^{+  0.4}_{-  0.3}$ & $152.0 \pm   8.5$ \\
\enddata 
\tablenotetext{a}{Measurements are in the quasar rest frame. If the lag is detected by {\tt JAVELIN}, we use the {\tt JAVELIN}-reported lag by default; otherwise, we use the {\tt CREAM}-reported lag.} 
\tablenotetext{b}{{Line widths are measured using Prepspec. The mean line widths are measured from the broad component only, and the rms line widths do not include the continuum.}}
\tablenotetext{c}{Virial products were converted to $M_{\rm BH}$ using $f$ = 4.47, as measured by \cite{Woo15}.} 
\tablenotetext{d}{From \cite{Shen15b}.} 
\label{Table:mbh_hb} 
\end{deluxetable*}  

\clearpage
\begin{deluxetable*}{lccrrrrrrr} 
\tablewidth{0pt} 
\tablecaption{Line Width, Virial Product, and $M_{\rm BH}$ Measurements for \Halpha} 
\tablehead{ 
\colhead{ } & 
\colhead{ } & 
\colhead{$\tau_{\rm final}$\tablenotemark{a}} & 
\colhead{$\sigma_{\rm line, mean}$} & 
\colhead{$\sigma_{\rm line, rms}$} & 
\colhead{FWHM$_{\rm mean}$} & 
\colhead{FWHM$_{\rm rms}$} & 
\colhead{VP} & 
\colhead{$M_{\rm BH}$\tablenotemark{b}} & 
\colhead{$\sigma_{*}$\tablenotemark{c}} \\ 
\colhead{RMID} & 
\colhead{$z$} & 
\colhead{(days)} & 
\colhead{(km s$^{-1}$)} & 
\colhead{(km s$^{-1}$)} & 
\colhead{(km s$^{-1}$)} & 
\colhead{(km s$^{-1}$)} & 
\colhead{(10$^7$ \Msun)} & 
\colhead{(10$^7$ \Msun)} & 
\colhead{(km s$^{-1}$)}  
}  
\startdata 
RM 017  &  0.456  &  $ 56.6^{+  7.3}_{- 15.1}$ & $ 4509 \pm    53$ & $ 4569 \pm    51$ & $ 4159 \pm    13$ & $ 5604 \pm    31$ & $ 23.1^{+  9.0}_{- 10.5}$ &  $103.1^{+ 40.2}_{- 46.8}$ & $191.4 \pm   3.7$ \\ 
RM 088  &  0.516  &  $ 54.8^{+  2.9}_{-  5.1}$ & $ 2449 \pm    27$ & $ 3320 \pm    26$ & $ 4451 \pm    32$ & $10290 \pm   142$ & $ 11.8^{+  4.4}_{-  4.5}$ &  $ 52.7^{+ 19.6}_{- 20.0}$ & $128.5 \pm  12.3$ \\ 
RM 160  &  0.359  &  $ 21.0^{+  1.4}_{-  2.8}$ & $ 1707 \pm     3$ & $ 1318 \pm    11$ & $ 3805 \pm    10$ & $ 3642 \pm    26$ & $  0.7^{+  0.3}_{-  0.3}$ &  $  3.2^{+  1.2}_{-  1.2}$ & \nodata \\ 
RM 191  &  0.442  &  $ 16.7^{+  4.1}_{-  5.5}$ & $  858 \pm     6$ & $  796 \pm    23$ & $ 2050 \pm    18$ & $ 1575 \pm    60$ & $  0.2^{+  0.1}_{-  0.1}$ &  $  0.9^{+  0.4}_{-  0.5}$ & $152.0 \pm   8.5$ \\ 
\smallskip 
RM 229  &  0.470  &  $ 22.1^{+  7.7}_{-  7.3}$ & $ 1528 \pm    10$ & $ 1738 \pm    31$ & $ 2271 \pm    34$ & $ 2103 \pm   365$ & $  1.3^{+  0.7}_{-  0.6}$ &  $  5.8^{+  2.9}_{-  2.9}$ & $130.2 \pm   8.7$ \\ 
RM 252  &  0.281  &  $ 10.1^{+  2.4}_{-  1.9}$ & $ 4300 \pm    26$ & $ 3384 \pm    71$ & $ 6574 \pm    69$ & $ 7868 \pm    66$ & $  2.3^{+  1.0}_{-  0.9}$ &  $ 10.1^{+  4.4}_{-  4.2}$ & $180.7 \pm   4.0$ \\ 
RM 272  &  0.263  &  $ 32.2^{+ 15.6}_{- 12.6}$ & $ 1408 \pm     1$ & $ 1298 \pm     8$ & $ 2436 \pm     3$ & $ 2632 \pm    28$ & $  1.1^{+  0.6}_{-  0.6}$ &  $  4.7^{+  2.9}_{-  2.5}$ & \nodata \\ 
RM 320  &  0.265  &  $ 20.2^{+ 10.5}_{-  9.3}$ & $ 1538 \pm     3$ & $ 1320 \pm    17$ & $ 3232 \pm    12$ & $ 2808 \pm    41$ & $  0.7^{+  0.4}_{-  0.4}$ &  $  3.1^{+  2.0}_{-  1.8}$ & $ 66.4 \pm   4.6$ \\
\enddata 
\tablenotetext{a}{Measurements are in the quasar rest frame. If the lag is detected by {\tt JAVELIN}, we use the {\tt JAVELIN}-reported lag by default; otherwise, we use the {\tt CREAM}-reported lag.} 
\tablenotetext{b}{Virial products were converted to $M_{\rm BH}$ using $f$ = 4.47, as measured by \cite{Woo15}.} 
\tablenotetext{c}{From \cite{Shen15b}.} 
\label{Table:mbh_ha} 
\end{deluxetable*}

\end{document}